# Investigating Mercury's Environment with the Two-Spacecraft BepiColombo Mission

A. Milillo[1] · M. Fujimoto[2] · G. Murakami[2] · J. Benkhoff[3] · J. Zender[3] · S. Aizawa[4] ·
M. Dósa[5] · L. Griton[4] · D. Heyner[6] · G. Ho[7] · S.M. Imber[8,9] · X. Jia[8] · T. Karlsson[10] ·
R.M. Killen[11] · M. Laurenza[1] · S.T. Lindsay[9] · S. McKenna-Lawlor[12] · A. Mura[1] ·
J.M. Raines[13] · D.A. Rothery[14] · N. André[4] · W. Baumjohann[15] · A. Berezhnoy[16,17] ·
P.A. Bourdin[15,18] · E.J. Bunce[9] · F. Califano[19] · J. Deca[20] · S. de la Fuente[21] · C. Dong[22] ·
C. Grava[23] · S. Fatemi[24] · P. Henri[25] · S.L. Ivanovski[26] · B.V. Jackson[27] · M. James[9] ·
E. Kallio[28] · Y. Kasaba[29] · E. Kilpua[30] · M. Kobayashi[31] · B. Langlais[32] · F. Leblanc[33] ·
C. Lhotka[15] · V. Mangano[1] · A. Martindale[9] · S. Massetti[1] · A. Masters[34] ·
M. Morooka[35] · Y. Narita[18] · J.S. Oliveira[3,36] · D. Odstrcil[11] · S. Orsini[1] ·
M.G. Pelizzo[37] · C. Plainaki[38] · F. Plaschke[15] · F. Sahraoui[39] · K. Seki[40] · J.A. Slavin[8] ·
R. Vainio[41] · P. Wurz[42] · S. Barabash[24] · C.M. Carr[43] · D. Delcourt[44] ·
K.-H. Glassmeier[6] · M. Grande[45] · M. Hirahara[46] · J. Huovelin[30] · O. Korablev[47] ·
H. Kojima[48] · H. Lichtenegger[15] · S. Livi[23] · A. Matsuoka[2] · R. Moissl[3] ·
M. Moncuquet[49] · K. Muinonen[30] · E. Quèmerais[50] · Y. Saito[2] · S. Yagitani[51] ·
I. Yoshikawa[52] · J.-E. Wahlund[35]

**Abstract** The ESA-JAXA BepiColombo mission will provide simultaneous measurements from two spacecraft, offering an unprecedented opportunity to investigate magnetospheric

✉ A. Milillo
anna.milillo@inaf.it

1   Institute of Space Astrophysics and Planetology, INAF, via del Fosso del Cavaliere 100, 00133, Rome, Italy

2   Institute of Space and Astronautical Science, Japan Aerospace Exploration Agency, Sagamihara, Kanagawa, Japan

3   Science and Operations Department, Directorate of Science, ESA/ESTEC, Noordwijk, The Netherlands

4   Institut de Recherche en Astrophysique et Planétologie, CNRS, CNES, Université de Toulouse, Toulouse, France

5   Department of Space Research and Space Technology, Wigner Research Centre for Physics, Budapest, Hungary

6   Institut für Geophysik und extraterrestrische Physik, Technische Universität Braunschweig, Braunschweig, Germany

7   The Johns Hopkins University Applied Physics Laboratory, Laurel, MD 20723, USA

and exospheric dynamics at Mercury as well as their interactions with the solar wind, radiation, and interplanetary dust. Many scientific instruments onboard the two spacecraft will be completely, or partially devoted to study the near-space environment of Mercury as well as the complex processes that govern it. Many issues remain unsolved even after the MESSENGER mission that ended in 2015. The specific orbits of the two spacecraft, MPO and Mio, and the comprehensive scientific payload allow a wider range of scientific questions to be addressed than those that could be achieved by the individual instruments acting alone, or by previous missions. These joint observations are of key importance because many phenomena in Mercury's environment are highly temporally and spatially variable. Examples of possible coordinated observations are described in this article, analysing the required geometrical conditions, pointing, resolutions and operation timing of different BepiColombo instruments sensors.

# 1 Introduction

Mercury's environment is a complex system where the magnetosphere and exosphere are inherently coupled, and interact with the interplanetary medium and the surface (e.g.: Milillo et al. 2005; Killen et al. 2007). The planet's close proximity to the Sun creates particularly strong external forcing conditions, involving extreme solar wind conditions and intense solar energetic particles and X-ray fluxes. Mercury possesses a weak, intrinsic, global magnetic

[8]  Department of Atmospheric, Oceanic and Space Sciences, University of Michigan, Ann Arbor, MI, USA

[9]  Space Research Centre, Department of Physics and Astronomy, University of Leicester, Leicester, UK

[10]  Space and Plasma Physics, School of Electrical Engineering and Computer Science, KTH Royal Institute of Technology, Stockholm, Sweden

[11]  Goddard Space Flight Centre, Maryland USA

[12]  Space Technology Ireland, Ltd., Maynooth, Kildare, Ireland

[13]  Department of Climate and Space Sciences and Engineering, Oceanic and Space Sciences, University of Michigan, Ann Arbor, MI, USA

[14]  School of Physical Sciences, The Open University, Milton Keynes, UK

[15]  Space Research Institute, Austrian Academy of Sciences, Graz, Austria

[16]  Sternberg Astronomical Institute, Moscow State University, Moscow, Russia

[17]  Institute of Physics, Kazan Federal University, Kazan, Russia

[18]  Institute of Physics, University of Graz, Graz, Austria

[19]  University of Pisa, Pisa, Italy

[20]  University of Colorado, Boulder, USA

[21]  ESA/ESAC, Villanueva de la Canada, Spain

[22]  Department of Astrophysical Sciences and Princeton Plasma Physics Laboratory, Princeton University, Princeton, NJ, USA

[23]  SWRI, San Antonio, TX, USA

field that supports a small magnetosphere which is populated by charged particles originating from the solar wind, from the planet's exosphere and from the surface (a comparison of Mercury's characteristics with Earth's is summarised in Table 1). On the other hand, the exosphere is continually refilled and eroded through a variety of chemical and physical processes acting both on the surface and in the planetary environment and which are driven by external conditions, like the Sun's irradiance and particles, and micrometeoroid precipitation toward the surface. These external conditions show a high variability along the eccentric orbit of Mercury (0.31–0.46 AU), so that, even though Mercury lacks seasons linked to its rotational axis inclination, it is generally assumed that the aphelion part of the orbit (True Anomaly Angle—TAA between 135° and 225°) is Winter, the perihelion (TAA between -45° and +45°) is Summer, while at TAA between 45° and 135° is Autumn and at TAA between 225° and 315° is Spring.

The first direct encounters with Mercury's environment comprised three flybys by the Mariner 10 spacecraft spanning 1974-75, allowing detection of Mercury's magnetic field, and observations of its exosphere and surface (Russell et al. 1988). In 2011, NASA's MESSENGER (Mercury Surface, Space Environment, Geochemistry and Ranging) mission (Solomon and Anderson 2018) was placed into a highly elliptical polar orbit around Mercury, carrying a suite of instruments designed to explore the fundamental characteristics of the planetary surface and environment. The mission concluded in 2015 with a low altitude campaign and finally impacted the planet.

Thanks to MESSENGER observations, we know that the Hermean magnetosphere is highly dynamic, with total reconfiguration taking place within a few minutes (Imber and Slavin 2017). The coupling of the interplanetary magnetic field and the solar wind with the planetary magnetosphere is much stronger than previously believed, owing to the almost-

[24]  Swedish Institute of Space Physics, Kiruna, Sweden

[25]  Laboratoire de Physique et Chimie de l'Environnement et de l'Espace, CNRS/Université d'Orléans/CNES, Orleans, France

[26]  Osservatorio Astronominco di Trieste, INAF, Trieste, Italy

[27]  University of California San Diego, La Jolla, USA

[28]  School of Electrical Engineering, Department of Electronics and Nanoengineering, Aalto University, Helsinki, Finland

[29]  Planetary Plasma and Atmospheric Research Center, Tohoku University, Sendai, Miyagi, Japan

[30]  Department of Physics, University of Helsinki, Helsinki, Finland

[31]  Planetary Exploration Research Center, Chiba Institute of Technology, Narashino, Japan

[32]  Laboratoire de Planétologie et Géodynamique, CNRS, Université de Nantes, Université d'Angers, Nantes, France

[33]  LATMOS/IPSL, CNRS, Sorbonne Université, Paris, France

[34]  The Blackett Laboratory, Imperial College London, London, UK

[35]  Swedish Institute of Space Physics (IRF), Uppsala, Sweden

[36]  CITEUC, Geophysical and Astronomical Observatory, University of Coimbra, Coimbra, Portugal

[37]  Institute for Photonics and Nanotechnologies, CNR, Padova, Italy

[38]  Italian Space Agency, Rome, Italy

[39]  Laboratoire de Physique des Plasmas, CNRS, Ecole Polytechnique, Sorbonne Université, Université Paris-Saclay, Observatoire de Paris, Meudon, France

**Table 1**  Mercury's and Earth's parameters

|  | Mercury | Earth |
| --- | --- | --- |
| Sun distance (AU) | 0.31–0.47 | 1 |
| Sidereal orbital period (Earth's day) | 87.97 | 365.26 |
| Inclination of orbit to solar equator (°) | 3.4 | 7.2 |
| Rotation period (Earth's day) | 58.6 | 1 |
| Inclination of rotation axis to orbit (°) | 0.034 | 23.4 |
| Mass ($10^{24}$ kg) | 0.33 | 5.97 |
| Radius (km) | 2440 | 6371 |
| Density (g cm-3) | 5.4 | 5.5 |
| Escape velocity (km/s) | 4.3 | 11.2 |
| Surface temperature (K) | 90–700 | 279 |
| Magnetic field moment | 195 nT $R_M^3$ (480 km Northward) | 31000 nT $R_E^3$ |
| Inclination of magnetic axis to rotation axis (°) | 0 | 11 |

continuous dayside magnetic reconnection (e.g. Slavin et al. 2012), as proved by the frequent observations of flux transfer events (FTE) (Imber et al. 2014; Leyser et al. 2017) (Fig. 1 upper panels). Slavin et al. (2014, 2019a) examined several MESSENGER passes during which extreme solar wind conditions both compressed the dayside magnetosphere due to high dynamic pressure, and eroded it due to extreme reconnection. The solar wind-planet interaction is further complicated by the existence of Mercury's large metallic core, within which induction currents are driven during these extreme events, acting in opposition to this compression/erosion (Jia et al. 2015, 2019; Dong et al. 2019). Eventually, the global current system within the magnetosphere and the surface is still an open question that can be solved only by multi-vantage point observations that allow discrimination between the inner and outer magnetic components.

[40]  Department of Earth and Planetary Science, Graduate School of Science, University of Tokyo, Tokyo, Japan

[41]  Space Research Laboratory, Department of Physics and Astronomy, University of Turku, Turku, Finland

[42]  Physics Institute, University of Bern, Bern, Switzerland

[43]  Department of Physics, Imperial College London, London UK

[44]  University of Orleans, Orleans, Paris, France

[45]  Institute of Mathematical and Physical Sciences, University of Wales, Aberystwyth, Wales, UK

[46]  Institute for Space-Earth Environmental Research, Nagoya University, Nagoya, Japan

[47]  IKI, Moscow, Russia

[48]  Research Institute for Sustainable Humanosphere, Kyoto University, Kyoto, Japan

[49]  LESIA, Observatoire de Paris, PSL Research University, CNRS, Sorbonne Université, UPMC, Université Paris Diderot, Sorbonne Paris Cité, Meudon, France

[50]  LATMOS, Université Versailles Saint-Quentin, Guyancourt, France

[51]  Graduate School of Natural Science and Technology, Kanazawa University, Kanazawa, Japan

[52]  Department of Complexity Science and Engineering, University of Tokyo, Tokyo, Japan

**Fig. 1** Upper left (noon-midnight meridional plane—NMP): Energetic planetary ions, ionized upstream of the magnetopause, are transported into the magnetosphere on newly reconnected field lines. Lower left (same plane): lower energy planetary ions are created by ionization inside the magnetopause. Upper right (same plane) Solar wind enters at the dayside via magnetopause reconnection. Lower right (Mercury's equatorial plane): Solar wind plasma and planetary ions mixing layers form at dusk low latitudes via Kelvin-Helmholtz waves. (Raines et al. 2015).

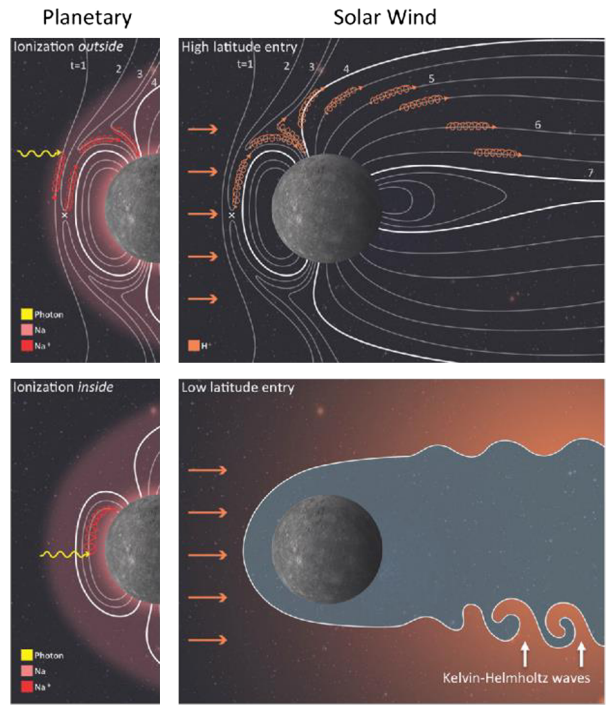

As predicted by Glassmeier and Espley (2006), MESSENGER observed Kelvin-Helmholtz (KH) instabilities especially at the dusk magnetopause (Fig. 1 lower left panel) that have been interpreted as the effect of the large gyroradius of heavy ions (Sundberg et al. 2012a). Nevertheless, due to lack of any wave – particle experiment and not having the possibility to reconstruct the full ion energy distribution, the energy and plasma transfer and excitation of ultra-low frequency (ULF) wave processes are not fully described. Furthermore, the role of planetary ions in magnetospheric processes is not well defined. In fact, we know that the planetary ions are circulating and responding to external conditions (Raines et al. 2014, 2015; Wurz and Blomberg 2001), but we do not know how these ions are generated and accelerated; also the mechanism able to energize electrons to energies up to hundreds of keV is still not identified, given the absence of radiation belts (Ho et al. 2016).

In the Hermean magnetotail, field-aligned currents have been quantified (Region-1 current; Anderson et al. 2014, 2018); dipolarization events have been observed in an Earth-like substorm manner by Mariner 10 and by MESSENGER (Sundberg et al. 2012b; Sun et al. 2015); the charged particles are convected toward the planet and the X-ray observations showed that electrons impact the planet's night-side as a result of dipolarization events (Lindsay et al. 2016); furthermore, during disturbed periods, flux ropes directed toward the far tail have been observed as signatures of tail reconnection events (Di Braccio et al. 2015a) (Fig. 2). However, Mercury has no ionosphere where the field-aligned currents close, nor is there evidence of a Region-2 current system. As hypothesised for the dayside magnetosphere, it is also supposed that the tail current system can close through Mercury's resistive crust and mantle at the conductive planetary core (Slavin et al. 2019b). Given that MESSENGER's eccentric orbit always had its apoherm at larger southern distances, it could not investigate the ion circulation in the far tail. The simultaneous measurements of particles and electromagnetic fields at high time and energy resolution from two positions along the tail by BepiColombo will allow a proper investigation. Also, the connection between precipitating particles and exosphere generation is still an open question, as well as the determination of the most effective surface release processes for each species (Killen et al. 2019; Gamborino et al. 2019) (Fig. 3).

We know from Earth-based observations that the Na exosphere is highly variable and seems to respond to solar conditions (Killen 2001; Mangano et al. 2013, 2015; Massetti et al. 2017). In fact, Interplanetary Coronal Mass Ejections (ICMEs; e.g., Kilpua et al. 2017) affect not only the magnetospheric configuration (Winslow et al. 2017; Slavin et al. 2014, 2019a), but also may allow ions to reach a large part of the dayside surface and consequently generate signatures in the sodium exosphere distributions (Orsini et al. 2018). However, the specific coupling mechanism is not yet fully understood mainly because MESSENGER was not able to simultaneously characterise the solar wind, the dayside reconnection rate, nor to make measurements of precipitating particles inside the magnetosphere, nor to perform plasma precipitation mapping via ion back-scattered neutral atom detection.

In contrast, the equatorial Na distributions seem dependent only on Mercury's orbital phase, according to space observations (Cassidy et al. 2015). Only simultaneous observations of the external conditions of solar wind, plasma precipitation, micrometeoroid, and exosphere distributions will allow a full understanding of the Na exospheric behaviour and help to solve the mystery of the highly volatile component in this close-to-star planet. Moreover, there is an indication of a strong correlation between the distribution of energetic refractory elements in the exosphere and the crossing of micrometeoroid streams (Killen and Hahn 2015), but MESSENGER had no dust monitor on board that might have been able to confirm this. We know from ground-based and from MESSENGER observations that the composition of the constituent particles in the Mercury's environment includes, besides H and He, Na and Na$^+$, K, Mg, Ca and Ca$^+$, Mn, Fe and Al (Broadfoot et al. 1974; Potter and Morgan 1985, 1986; Bida et al. 2000; McClintock et al. 2008; Bida and Killen 2017), while unexpectedly, no signature of oxygen atoms was detected by the Mercury Atmospheric and Surface Composition Spectrometer (MASCS; McClintock and Lankton 2007). However, the mass resolution of the Fast Imaging Plasma Spectrometer (FIPS; Raines et al. 2011) was too low to discriminate between individual ion species, while atom groups including oxygen could not be detected by MASCS. Mass spectrometers with high mass resolution would allow the detection and characterisation of the majority of the constituent of the exosphere and planetary ions including molecules and atom groups that would provide an important information for describing the surface release processes, for explaining the fate of oxygen, and ultimately for tracing the planet's evolutionary history.

In summary, in the 1980s, the analysis of the data of the three Mariner-10 fly-bys revealed unexpected features of Mercury's environment including the intrinsic magnetic field and the presence of high-energy electron burst events. With the new millennium, the MESSENGER mission, thanks also to the exosphere ground-based observations, have greatly improved our knowledge of the complex Hermean environment. However, both missions left the planet with new intriguing questions. In Sect. 2, we summarise the main findings about the Hermean environment and the still unsolved points.

The ESA-JAXA BepiColombo mission is expected to provide a tremendous improvement in the knowledge of the functioning of Mercury's environment, and solve the numerous questions that are still open after previous space missions together with ground-based observations. In fact, this technologically advanced and optimally designed mission exhibits

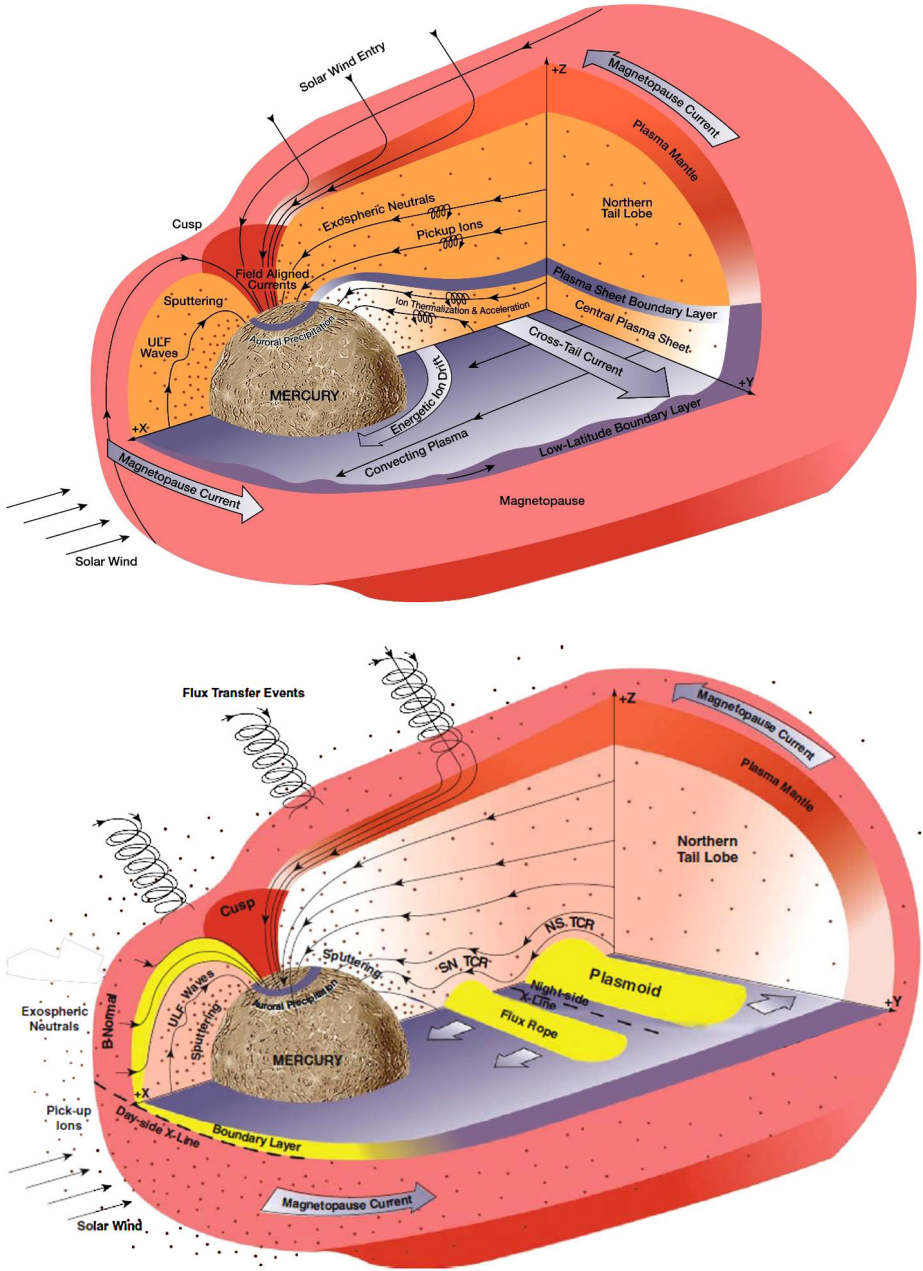

**Fig. 2** Mercury's magnetosphere schematic view: (**a**) in its ground state during northward IMF: minimal magnetic field normal to the magnetopause, less reconnection, weak tail magnetic field, thick plasma sheet, no near-Mercury reconnection X-lines, and well-developed low-latitude boundary layer (LLBL). (**b**) in active period during southward IMF: more reconnection, substorm onset in a highly stressed magnetosphere with large magnetic fields normal to the magnetopause, a strongly loaded tail, a thinned plasma sheet, multiple near-Mercury X-lines, and plasmoids. (from Slavin et al. 2019b)

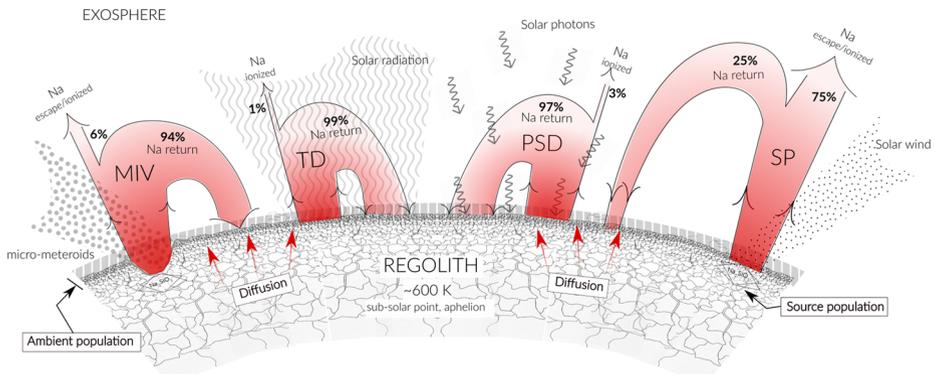

**Fig. 3** Scheme illustrating the different released fluxes of Na due to the different release mechanisms. The illustration is not to scale. (Gamborino et al. 2019)

all the capabilities to accomplish the requirements mentioned above. It allows simultaneous two-point measurements thanks to the two spacecraft, MPO and Mio, with optimal orbits for exploring external, close-to-planet, far-tail and flank conditions (Milillo et al. 2010). BepiColombo, unlike MESSENGER, has a dedicated full plasma instruments package in the Mio spacecraft that offers a unique opportunity to study in details the magnetosphere of Mercury. Among the instruments devoted to the study of the Mercury's environment, Bepi-Colombo includes sensors and experiments that never have been operated at the innermost planet, like the plasma wave experiment Mio/PWI, the dust monitor Mio/MDM, the neutral mass spectrometer MPO/SERENA-STROFIO and Na imager Mio/MSASI, two Energetic Neutral Atom (ENA) imagers MPO/SERENA-ELENA and Mio/MPPE-ENA. Furthermore, many of the BepiColombo instrument types previously flown on Mariner-10 and MESSEN-GER (i.e.: magnetometers MPO/MAG and Mio/MGF, charged particle detectors Mio/MPPE and MPO/SERENA, UV spectrometer MPO/PHEBUS, X-rays spectrometers MPO/MIXS and SIXS) have significantly improved performances in spatial coverage, energy, mass and time resolutions. In other words, BepiColombo will offer an unprecedented opportunity to investigate magnetospheric and exospheric dynamics in the deepest level ever reached at Mercury as well as their interactions with solar wind and radiation, and interplanetary dust. In Sect. 3, the main characteristics of the BepiColombo composite mission and of instruments for the environment are briefly described.

In Sect. 4, some possible joint investigations performed by coordinated measurements of different instruments on board of the two spacecraft are suggested. These are intended as examples of the potentialities of the BepiColombo mission for the study of the coupled magnetosphere-exosphere-surface-interior system of Mercury. Summary and conclusions are given in Sect. 5.

## 2 Findings and Open Questions on the Hermean Environment

Thanks to Mariner-10 three fly-bys, the MESSENGER mission and the ground-based observations of the exosphere, we now have some grasp on the complexity of the Hermean environment. Starting from the space environment at Mercury's orbit, proceeding with the magnetosphere, the exosphere and finally the surface relevant characteristics, in this section, we provide a summary of the current state of knowledge of Mercury's environment, along with a discussion of the related unanswered key questions.

**Table 2** Solar wind parameters at Mercury's and Earth's orbit

| Parameter | Mercury | Earth |
|---|---|---|
| Sun distance | 0.31–0.47 AU | 1 AU |
| Solar wind speed (km/s) | 250–650 | 320–710 |
| Solar wind density (cm$^{-3}$) | 15–105 | 3.2–20 |
| Proton temperature ($10^4$ K) | 13–17 | 8 |
| Interplanetary magnetic field (nT) | $31 \pm 11$ | $\sim$6 |
| Parker's spiral angle (°) | $\sim$20 | $\sim$45 |
| Alfvénic-Mach number | 2–5 | 6–11 |

## 2.1 Solar Wind, Radiation Environment and Dust at Mercury Orbit

It is necessary to understand the background solar wind, radiation environment and dust conditions at Mercury to interpret future measurements and identify specific space weather characteristics. Observations taken by the Helios 1 and 2 and MESSENGER missions have characterised the interplanetary conditions at the orbit of Mercury (Marsch et al. 1982; Pilipp et al. 1987; Sarantos et al. 2007; James et al. 2017; Korth et al. 2011).

The key solar wind parameters that influence planetary space weather are the flow speed ($v$), density ($n$), proton and electron temperatures ($T_p$, $T_e$) and interplanetary magnetic field (IMF) strength and direction and parameters derived from them, like plasma and magnetic pressures, Alfvénic/sonic numbers and the plasma beta, $\beta$, (e.g., Pulkkinen 2007; Lilensten et al. 2014; Plainaki et al. 2016). The solar wind speed does not change significantly with radial distance from the Sun, its average value being 430 km/s, however it shows a significant variability (peaks of 800 km/s). The solar wind density and the strength of the IMF decrease with the square of heliocentric distance, so that on average the density and IMF strength at Mercury's orbit are 5–10 times larger than that at the Earth orbit (see Table 2) (Burlaga 2001; Slavin and Holzer 1981). The Parker spiral at Mercury orbit forms an angle of about 20° with the solar wind flow direction, which implies a change of the relative magnitude of the IMF components with respect to the near-Earth conditions (the angle at the Earth's orbit is $\sim$45°). The Alfvénic Mach number ($M_A = V_{sw}/V_A$, where $V_A$ is the Alfvén speed) at Mercury is about 2–5, while it is between 6–11 at the Earth (Winterhalter et al. 1985) and consequently also $\beta$, the ratio between the plasma and the magnetic pressures, is lower at Mercury, ranging between 0.5 and 0.9 against 1.7 at the Earth (Slavin and Holzer 1981). The average parameters are also slightly variable depending on the solar cycle phase (e.g.: Schwenn 1990; Korth et al. 2011). A summary of the typical solar wind parameters at Mercury and at the Earth is given in Table 2.

Corotating Interaction Regions (CIRs), High Speed Streams (HSSs) and ICMEs are commonly observed at Mercury. During these events the plasma conditions in the solar wind are known to vary significantly from the average.

HSSs are a domain of solar wind plasma flowing at a higher speed than usual, typically reaching a speed of 700 to 800 km/s. They are characterised by relatively weak IMF, with rapidly changing direction due to Alfvenic fluctuations, and low density. They are considered to originate in the coronal holes on the solar surface in which the magnetic field forms an open-field structure. Very high $M_A$ might be observed at Mercury's orbit during HSSs at solar maximum (Baumjohann et al. 2006). CIRs (e.g., Pizzo 1991; Richardson 2018) are flow structures evolving in the background solar wind due to a velocity difference between adjacent plasma streams, e.g. slow solar wind and HSSs. A stream interface forms between the two different plasma regimes and develops to a well defined structure near the orbit of Earth. At the orbit of Mercury, CIRs are typically not yet evolved (Dósa and Erdős 2017) or are less pronounced (Schwenn 1990). As CIRs evolve radially outward, compression and shear between the two streams increases. These two factors give rise to fluctuations that are superposed upon Alfvénic fluctuations generated close to the Sun. This means that inner, "younger" regions of interplanetary plasma tend to carry signatures of solar origin and their investigation can provide insight into solar processes. Measurements of magnetic field fluctuations at low frequencies can help to constrain different models of solar wind heating mechanisms and acceleration by low frequency waves (e.g. Hollweg and Isenberg 2002; Dong and Paty 2011; Dong 2014; Suzuki 2002).

ICMEs (e.g., Sheeley et al. 1985; Gopalswamy 2006; Kilpua et al. 2017) are macro-scale interplanetary structures related to Coronal Mass Ejections (CMEs) characterised generally by a higher fraction of heavy multi-charged ions (Galvin 1997; Richardson and Cane 2004). Their integral part is a magnetic flux rope and if sufficiently faster than the preceding solar wind, ICMEs have leading shocks and turbulent sheath regions ahead. Winslow et al. (2015) studied 61 ICMEs detected by MESSENGER and found high magnetic field intensity and fast mean velocity (86.2 nT, and 706 km/s, respectively). Good et al. (2015) analysed the radial evolution of a magnetic cloud ICME, using data from MESSENGER and Solar Terrestrial Relations Observatory (STEREO)-B, and found evidence that the structure was clearly expanding, with a radius increasing by about a factor of two between Mercury's and Earth's orbits. Unlike HSSs, CIRs and the ICME sheaths, ICMEs generally present very low $M_A$ at the orbit of Mercury. Study of ICME propagation has been carried out in the past by using Helios1, Helios 2 and IMP data (Burlaga et al. 1980). The in situ observations of the interplanetary conditions at Mercury's orbit by BepiColombo/Mio instrumentation coupled with observations at 1 AU or at different distances from the Sun, performed by other space missions like Solar Orbiter (Müller et al. 2013) or Parker Solar Probe (Fox et al. 2016), could be compared to the results of propagation disturbance models at varying heliospheric distances or used to constrain them (e.g., Möstl et al. 2018) and for improving the knowledge of the acceleration mechanisms. In Sect. 4.1, coordinated measurements by BepiColombo and other missions coupled to possible models/tools for the interpretations are suggested.

Galactic Cosmic Rays (GCRs) are a homogeneous, nearly isotropic background of high-energy charged particles (mostly protons) with an energy reaching GeV to even $10^{24}$ eV, originating outside the Solar System, and constituting an important component of the particle radiation environment at Mercury. They continuously bombard Mercury's surface, generating cascades of secondary particles, including neutrons and gamma rays, providing a diagnostic of the Mercury surface composition (e.g., Goldsten et al. 2007). MESSENGER observations suggested that GCR protons are a potential energy source to stimulate organic synthesis at Mercury's poles, where wide water ice deposits are thought to be present in permanently shadowed regions, which may contain organics (e.g., Lawrence et al. 2013; Paige et al. 2013). To characterise the Hermean radiation environment and better understand this phenomenon, an accurate evaluation of the GCR flux at Mercury's orbit is needed, which so far has been possible only through modelling of the GCR propagation in the heliosphere (e.g., Potgieter 2013). BepiColombo will be able to monitor the GCR radiation environment and to estimate its intensity and modulation features (see Sect. 4.1).

The interplanetary medium is also populated by dust grains. Three major populations of the interplanetary dust have been identified in the inner solar system (0.3 to 1.0 AU) by previous in-situ dust observations, the Pioneer 8 and 9 and Helios dust experiments (e.g., Grün et al. 2001). Particles of one population have low-eccentricity orbits about the Sun and are related to particles originating in the asteroid belt, while particles of the second population have highly eccentric orbits and are allegedly emitted from short-period comets (Dermott et al. 2001; Jackson and Zook 1992). In situ measurements of those particles revealed grains with size from 100 down to 1 micrometer at an impact speed of 10 km/s. For the interplanetary grains in that size range, the dynamics is primarily affected by the gravitational force of the Sun, $F_{gr}$, and the solar radiation pressure, $F_{rad}$, the ratio $\beta = F_{rad}/F_{gr}$ being close to unity. Due to the component of the radiation force tangential to a grain's orbit, called Poynting-Robertson Light drag (e.g., Dermott et al. 2001), micron-sized particles spiral down toward the Sun. The third population identified in the inner solar system, called "$\beta$ meteoroids", composed of small particles in size range between tens of nanometer and 0.1 micrometer and detected to arrive from the solar direction (Iglseder et al. 1996). Due to such small size, those dust particles are accelerated radially outwards by the solar radiation force against solar gravity and finally they could reach escape velocity, therefore, having hyperbolic orbits they exit the solar system (Zook and Berg 1975; Wehry and Mann 1999). In addition to these dust populations, the presence of a circumsolar ring near Mercury's orbit was recently recognized by remote observation (Stenborg et al. 2018).

Dust grains are charged due to UV radiation or collision with charged particles; hence, they are subject to the Lorentz force that for sub-micron dust grains in the inner solar system becomes more important than the other forces, such as gravity and radiation pressure (Leinert and Grün 1990). The electric potential depends on the density and temperature of the surrounding plasma as well as the photoelectron intensity due to solar radiation. Therefore, the charge number on the grain, proportional to the dust size and potential, can be dynamic.

Using in situ measurements Meyer-Vernet et al. (2009) revealed the presence of anti-sunward directed nanograins near the Earth's orbit. Interestingly, the distribution of the nanograins at 1 AU is highly variable, where periods of high- and zero-impact rates alternate with a period of about 6 months (Zaslavsky et al. 2012). This dust structure could be due to the complex dynamics of the charged grains in a non-uniform solar wind structure (Juház and Horányi 2013). Consequently, the electromagnetic properties in the solar wind are important to understand the dynamics of the sub-micron and micrometer sized dust.

These dust grains/micro-meteoroids impact Mercury's surface along its orbit. Mercury has an inclined orbit, and since Mercury is away from the ecliptic plane at aphelion, it is expected that close to the aphelion phase the flux of meteoroids impinging on the surface of Mercury and the flux of ejecta particles will both decrease (Kameda et al. 2009). The impacting rate and the local time asymmetries are poorly characterised (Pokorný et al. 2017), but a clear relation with a comet stream crossing has been observed in the exosphere composition by MESSENGER/MASCS (Killen and Hahn 2015) (See Sects. 2.6 and 2.8). The interplanetary dust grains will be detected and characterised by BepiColombo Mio instrumentation, for the first time at the Mercury environment and related to the exospheric distribution and composition (see Sect. 4.9).

## 2.2 What Is the Magnetosphere Configuration, Its Relation with the Planet's Interior Structure and Its Response to Solar Activity?

The Hermean dipole moment is relatively weak ($m_M = 195$ nT $\cdot R_M^3$; almost perfectly aligned with the rotational axis as derived by the average MESSENGER measurements of Anderson et al. 2012). The planet is engulfed by the inner heliosphere solar wind with relatively intense dynamic pressure ($p_{sw,M} = \frac{1}{2}\rho_M v_{sw}^2 \approx 10$ nPa). In comparison, the terrestrial values are very different ($m_E \approx 31000$ nT $\cdot R_E^3$; $p_{sw,E} = \frac{1}{2}\rho_E v_{sw}^2 \approx 0.5$ nPa). Nonetheless, Mercury's intrinsic magnetic field interaction with the solar wind results in formation of a proper planetary magnetosphere, which is unique in the Solar System, being the only one of the same length scale as the planet itself. The structure of the magnetosphere resembles the terrestrial one, but differs in details. On the dayside, the planetary magnetic field is compressed by the solar wind flow, while on the night-side the magnetic field lines become stretched and elongated away from the planet and form two lobe regions in the tail separated by the current sheet. The outer boundary of Mercury's magnetosphere towards the magnetosheath is the magnetopause, whereas the inner boundary is the surface itself. Due to the weak magnetic field of the planet and the high dynamic pressure in the solar wind, only a small magnetosphere is created. The average sub-solar magnetopause distance is only 1.41 $R_M$ from the planet center (Korth et al. 2017), while at the Earth it is about 10 $R_E$. The relatively strong interior quadrupole moment with respect to the dipole causes a northward shift of the equatorial magnetosphere by 0.196 $R_M$ (Anderson et al. 2012; Johnson et al. 2012; Wicht and Heyner 2014). This dipole offset has been the result of an analysis of the MESSENGER magnetic equator crossing done in the range $3150 \leq \rho_z \leq 3720$ km (with $\rho_z$ as distance to the planetary rotation axis). Other analysis methods yielded different values of the dipole offsets. Thébault et al. (2018) reports an offset of 0.27 $R_M$.

Mercury's small magnetosphere, therefore, controls, guides, and accelerates the solar wind plasma and solar energetic particles such that charged particles (>keV) precipitation can occur with enhanced intensity focused at particular locations on the surface (see Sect. 2.4). This is in contrast to the Moon or asteroids where one side of the object is bathed by unfocused solar wind, which, apart from solar eruptions, usually has lower energies (about 1 keV/nucleon) (Kallio et al. 2008). In addition, the low $M_A$ of the solar wind causes Mercury's bow shock and magnetopause boundary to vary dynamically over short timescales.

As Mercury does not possess an ionosphere, the planet body is directly subject to magnetospheric variations. Changes in the external magnetic field (e.g. from the magnetospheric dynamics) drive currents within the electrical conducting interior of the planet (e.g.: Janhunen and Kallio 2004). As the electromagnetic skin depth $\delta$, i.e. the characteristic depth to which a changing magnetic field penetrates a conductor, depends on the frequency of variations and the conductivity $\sigma$ of the material as $\delta = \sqrt{\frac{(2)}{\omega \mu_0 \sigma}}$, one has to consider the frequency band of the variation as well as the conductivity structure of the planet. Using models for the closure of field-aligned currents as observed by MESSENGER through the planet, Anderson et al. (2018) estimated the planetary conductivity structure. There the conductivity exponentially rises with depth from the crust/mantle ($\sigma \approx 10^{-8}$ S/m) to the highly conducting core-mantle boundary ($\sigma \approx 10^6$ S/m) at $r \approx$ 2000 km from the planet center (Hauck et al. 2013; Johnson et al. 2016). They estimated that up to 90% of the total current might close in this manner (Anderson et al. 2018). Short time variations penetrate only the upper planetary layers whereas long time variations may penetrate to the core causing induction currents (e.g., Hood and Schubert 1979; Suess and Goldstein 1979; Glassmeier et al. 2007a). The effects of the induction currents on the large-scale configuration of Mercury's magnetosphere have been inferred from the MESSENGER data for cases of extreme (Slavin et al. 2019a), strong (Slavin et al. 2014; Jia et al. 2019) and modest (Zhong et al. 2015a, 2015b; Johnson et al. 2016) variations in solar wind pressure. Global simulations that self-consistently model the induction effects (Jia et al. 2015, 2019; Dong et al. 2019) have clearly demonstrated that the shielding effect of induction and reconnection-driven erosion compete against each other for dominance in controlling the large-scale structure of Mercury's magnetosphere (Fig. 4). During extreme events, intense reconnection at the dayside magnetopause is expected and this reduces the magnetopause stand-off distance (e.g., Slavin and Holzer 1979; Jia et al. 2019; Slavin et al. 2019a) from 1.4 $R_M$ down to 1.03 $R_M$: larger/smaller magnetopause stand-off distances are correlated with lower/higher reconnection rates. It is vital to understand the magnitude of such induction currents, as they temporarily change the magnetic dipole moment of the planet, acting to prevent the solar wind from directly impacting the planetary surface (Heyner et al. 2016). In fact, the effective magnetic moment inferred by the magnetopause stand-off distance and plasma pressure is not univocally fixed but it ranges between 170 and 250 nT–$R_M^3$ (Jia et al. 2019) depending on external condition and by magnetic pressure time gradients.

Important science questions that BepiColombo can answer are: how do the currents circulate inside the planetary crust? to what extent does the planetary field shield the planetary surface from direct impact of particles from the solar wind on the dayside and from the central plasma sheet on the night side? is the shielding effective only during the largest induction events, or always effective except during the most intense reconnection events, or at some intermediate point between these two extremes? The planned orbits for the two spacecraft will enable Mio to acquire direct measurements of the upstream solar wind while at the same time MPO will monitor the space environment close to the planet. Such a conjunction between the two spacecraft is ideal for studying Mercury's planetary response to the external solar wind forcing (see Sect. 4.3)

Not only do the induction currents produce dayside magnetosphere reconfiguration but, simultaneously, the nightside current systems are also significantly altered (Fig. 4 upper panel) (see Sect. 2.4). This delicate interplay between induction and reconnection, proposed by Slavin and Holzer (1979), was estimated by Heyner et al. (2016). Johnson et al. (2016) showed that the 88-day-variation in the magnetosphere due to the planetary orbit around the Sun changes the dipole moment of the planet (by about 4%) by driving induction currents deep inside the planet. Thereby, measurements of the variation of Mercury's magnetospheric structure can be used to constrain its core-mantle boundary independently from geodetic measurements. By studying correlated periodic temporal variations, of external and induced origins, Wardinski et al. (2019) estimated the size of the electrically conductive core to be 2060 km, slightly above previous geodetic estimates. Variations on geological time scales may actually penetrate inside the core and give rise to a negative magnetospheric feedback on the interior dynamo (Glassmeier et al. 2007b; Heyner et al. 2011). The dual probe BepiColombo mission is highly suited to further study the relation between day- and nightside processes in particular during configuration with one spacecraft at the dayside and the other in the magnetotail (see Sect. 4.7). Moreover, the north-south symmetry of the BepiColombo orbits will allow a characterisation of the southern hemisphere environment, which was not well covered by MESSENGER. Furthermore, any long-term variations in the magnetospheric field may be used to sound the electrical conductivity structure of the planet in a magnetotelluric fashion.

Last but not least, the final phase of the MESSENGER mission enabled the discovery of crustal magnetic anomalies in the northern hemisphere (Hood et al. 2018). Their analysis may provide important information about the temporal variation of the planetary magnetic field in the past (Oliveira et al. 2019). However, if the southern hemisphere magnetic anomalies are similar to those of the northern hemisphere, where the magnetic field arising from the known anomalies is at maximum 8 nT at 40 km (Hood 2016; Hood et al. 2018), their effect should be negligible for magnetospheric dynamics. Possible deviation of charged particles at the surface by local magnetic fields would be recognised by BepiColombo as ion back-scattering intensification at the interface between the micro-magnetosphere and its internal cavity, as it has been observed in the case on Mars (Hara et al. 2018) and the Moon (Saito et al. 2008; Deca et al. 2015; Poppe et al. 2017).

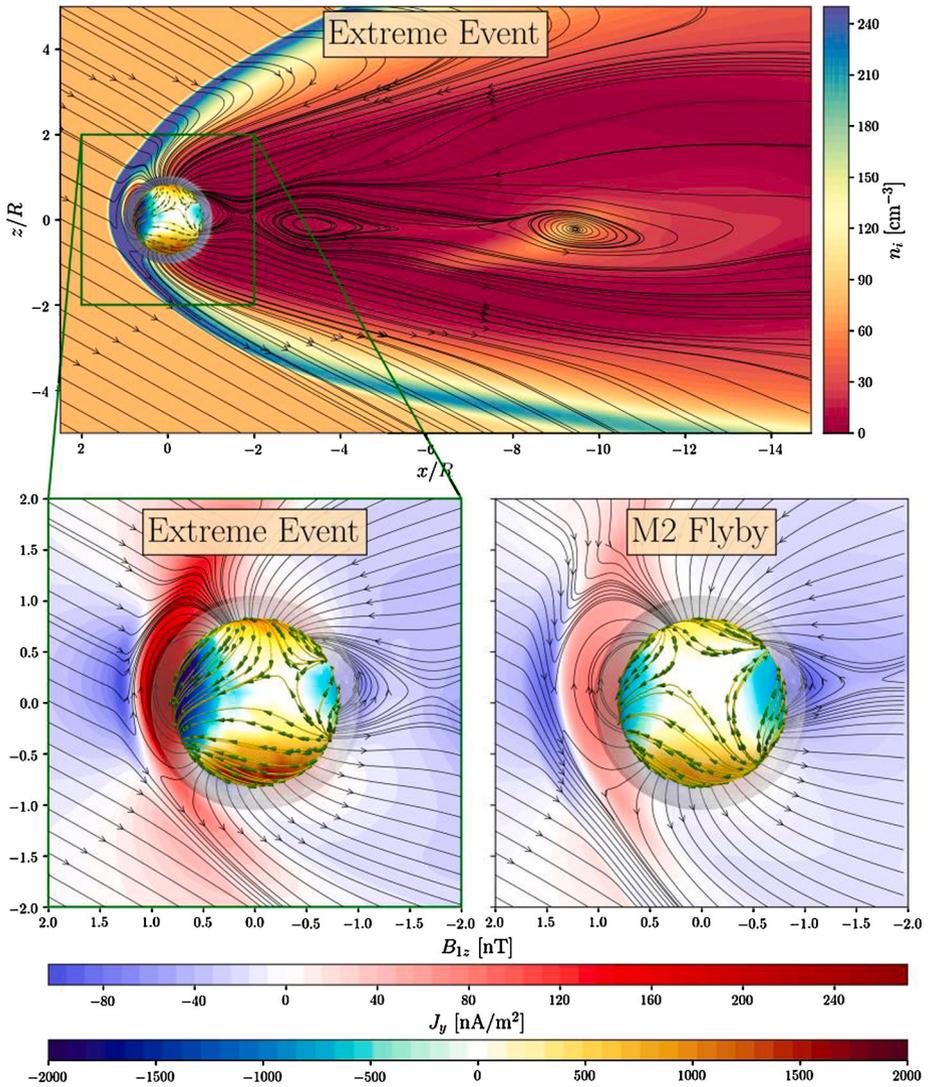

**Fig. 4** Mercury's magnetosphere in x-z (meridian) plane during an extreme event from the calculation of Dong et al. (2019). Plasmoids (or flux ropes) are formed in Mercury's magnetotail. The background color contours in the upper panel show the ion density per cubic centimeter. The lower left panel shows the zoomed-in subdomain where color contours in x-z plane represent the perturbation magnetic field B1z (in nanotesla) and the color contours on the conducting core surface are the induction current Jy (in nanoamperes per square meter). Note that the streamlines of core surface currents are illustrated by the yellow curves with green arrows wrapping around the core. Compared with the lower right panel of M2, the B1z and the induction current Jy from the extreme event are much stronger

## 2.3 How Does the Solar Wind Mix with Mercury's Magnetosphere?

The pristine solar wind does not directly interact with the Hermean magnetosphere. Instead, it is modified by processes in the upstream foreshock, at the bow shock, and in the magne-tosheath before encountering the magnetopause. At the bow shock, the solar wind plasma is decelerated and heated from super-magnetosonic to sub-magnetosonic speeds, enabling it to flow around the obstacle that the Hermean magnetosphere constitutes (e.g., Anderson et al. 2010, 2011). As the IMF cone angle (the angle between the IMF and the Mercury-Sun-line) is typically ∼20° (Table 2), the subsolar bow shock is most often quasi-parallel (e.g., Slavin and Holzer 1981).

Upstream of the quasi-parallel shock, a foreshock region can be found that is magnetically connected to the shock, into which shock-reflected (so-called back-streaming) particles are able to travel along the IMF (Jarvinen et al. 2019). Those particles interact with the solar wind, generating waves and steepened magnetic structures (e.g., Burgess et al. 2005; Jarvinen et al. 2019). These waves and structures are convected with the solar wind stream back to the quasi-parallel shock. Hence, the regions upstream and downstream of that shock are generally more variable in comparison to the quasi-perpendicular shock and adjacent regions (e.g.: Le et al. 2013 Eastwood et al. 2005; Sundberg et al. 2013, 2015; Karlsson et al. 2016; Jarvinen et al. 2019).

The terrestrial quasi-parallel bow shock is highly structured and allows for high-speed jets of solar wind plasma to regularly form, penetrate the magnetosheath, and impact onto the magnetopause (e.g., Hietala et al. 2009, 2012; Plaschke et al. 2013a, 2013b, 2018). Signatures of high-speed jets have not yet been found in the Hermean magnetosheath (Karlsson et al. 2016), however, structures similar to hot flow anomalies have been identified near Mercury (Uritsky et al. 2014).

Differences are also apparent with respect to the quasi-perpendicular side of the bow shock and the corresponding magnetosheath, where ion cyclotron and mirror mode waves can originate from anisotropic particle distributions (e.g. Gary et al. 1993). At Earth, both modes exist, while at Mercury, only ion cyclotron waves have been observed (Sundberg et al. 2015). Mirror modes have only been predicted in simulations (Herčík et al. 2013). Their growth in the dayside magnetosheath may be inhibited by the limited size of the region and by the low plasma $\beta$ (Gershman et al. 2013).

Mio observations are expected to shed light on the existence and basic properties of several foreshock and magnetosheath phenomena, including foreshock cavities, bubbles, hot flow anomalies, jets, and mirror mode waves, due to the optimised orbit and advanced plasma instrumentation with respect to MESSENGER. In addition, Mio and MPO dayside conjunctions will allow, for the first time, simultaneous observations near Mercury and in the upstream foreshock, shock, or magnetosheath regions. This will make it possible to study the impact of transient phenomena emerging in these regions of the Hermean magnetosphere.

Two-point measurements in the magnetosheath will also give information on how the turbulence develops downstream of the bow shock, which will give an interesting comparison to the situation at Earth. Turbulence is probably the best example in plasma physics of multi-scale, nonlinear dynamics connecting fluid and kinetic plasma regimes and involving the development of many different phenomena spreading the energy all over many decades of wave numbers. To date, the near-Earth environment and the solar wind represented the best laboratory for the study of plasma turbulence (Bruno and Carbone 2013 and references therein) providing access to measurements that would not be possible in laboratories. Turbulent processes were observed in MESSENGER magnetic field data (Uritsky et al. 2011), especially at kinetic scales. Thanks to the Mio full plasma suite and to the MPO instruments for the space plasma observations, BepiColombo will offer the opportunity not only to conduct thorough turbulent studies, but also the great opportunity, from a physical point of view, to access the physical parameters and, therefore, the plasma regimes that are not available in the terrestrial magnetosphere and nearby solar wind. In particular, we will have access to low beta regimes inside the Mercury magnetosphere and to a fully kinetic turbulence lacking the large-scale MHD component typical of the Earth's magnetosheath. Last but not least, during operations of BepiColombo at Mercury and in the nearby solar wind, combined analysis with Solar Orbiter and Parker Solar Probe are expected to be a great opportunity to build a more complete view on the inner solar wind turbulence properties at different distances from the Sun.

The solar wind flowing around the Hermean magnetosphere, producing turbulence, is a driver of both magnetic and plasma fluid instabilities eventually producing an efficient mixing of the two plasmas. In this context, magnetic reconnection plays a key role by ultimately allowing for the entering of solar wind plasma into the magnetosphere, and thus a net momentum transport across the magnetopause. Nevertheless, it is not yet fully understood how dayside reconnection is triggered at the sub-solar point of Mercury's magnetopause. Based on observations at the Earth, magnetic reconnection between the southward oriented IMF and the planetary magnetic field is the most effective plasma mixing process. However, analysis of MESSENGER data demonstrated that reconnection at Mercury is significantly more intense than at the Earth (e.g. Slavin et al. 2009, 2012, 2014). Di Braccio et al. (2013) reported that the reconnection rate in the subsolar region of the magnetopause is independent of the IMF orientation, attributed to the influence of low-$\beta$ plasma depletion layers (Gershman et al. 2013), however a larger statistical study found that reconnection-related signatures were observed at a significantly higher rate during southward IMF intervals, and concluded that the relationship between clock angle and reconnection rate is akin to that observed at the Earth (Leyser et al. 2017).

BepiColombo will consistently provide very good estimates of the plasma pressure values, enabling more comprehensive studies of this phenomenon. A much better understanding of the dayside reconnection processes at Mercury is crucial, being the dominant process allowing for the solar wind plasma to enter the magnetosphere. Frequent reconnection made the measurement of large amplitude Flux Transfer Events (FTEs) (Slavin et al. 2012; Imber et al. 2014; Leyser et al. 2017) by MESSENGER a common occurrence. Solar wind particles reaching the cusps are eventually partially mirrored in the strengthening field or impacting the surface there, as observed by MESSENGER (Winslow et al. 2014; Raines et al. 2014). Poh et al. (2016) observed isolated, small-scale magnetic field depressions in the dayside magnetosphere, known as cusp filaments, thought to be the low latitude extent of FTEs (Fig. 5). Since this particle bombardment at the cusp regions is on going over geological time scales, the surface material may actually be darkened in certain spectral bands (see also Sect. 2.7 and Rothery et al. 2020, this issue). The cusp location depends on the Hermean heliocentric distance as well as the IMF direction. The northern cusp region has been readily identified by analysing the magnetic field fluctuations and its anisotropy related to the reconnection (He et al. 2017). The BepiColombo two-spacecraft configuration in the cusp region will offer an optimal opportunity for a detailed analysis of FTEs, filaments, and plasma entering the magnetosphere in both hemispheres (see Sect. 4.5).

Magnetopause reconnection is not limited to the dayside. Müller et al. (2012) showed in a simulation how reconnection at the equatorial dawn flank allows magnetosheath plasma to enter the magnetosphere and contribute to a partial ring current plasma. Di Braccio et al. (2015a) provided the first observations of the plasma mantle, a region in the near-tail where plasma is able to cross the magnetopause along open field lines. A subsequent statistical analysis by Jasinski et al. (2017) demonstrated that the mantle was more likely to be observed during southward IMF, and (due to the observations being entirely in the southern hemisphere), during periods of negative $B_X$. The BepiColombo orbit will allow an in-situ analysis of such phenomena in both hemispheres (see Sect. 4.2).

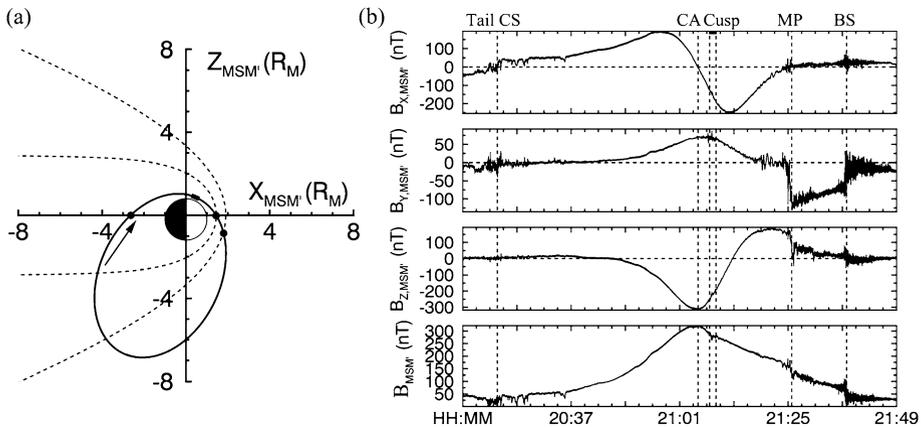

**Fig. 5** (**a**) Example of a MESSENGER orbit (black solid line) on 26 August 2011 projected onto the X-Z plane in aberrated Mercury solar magnetic (MSM') coordinates during a period without filamentary activities in the cusp. The model bow shock (*BS*) and magnetopause (*MP*) from Winslow et al. (2013) (marked by the two dots at the dayside magnetosphere) are shown in dotted lines; the Sun is to the right. The thick portion of the orbit represents the cusp region, and the dot at the nightside magnetosphere represents the magnetotail current sheet (*CS*) crossing. The arrow denotes the spacecraft trajectory. (**b**) Full-resolution magnetic field measurements (top to bottom, *X*, *Y*, and *Z* components and field magnitude) acquired along the orbit shown in Fig. 1a. The vertical dashed lines mark the boundary crossings shown in panel (**a**). CA denotes the closest approach, and all times are in UTC. (Poh et al. 2016)

Apart from reconnection processes, instabilities (e.g. KH, mirror and firehose instabilities) also play an important role in the mixing process and can be associated with different kinds of waves. At Mercury, KH instabilities were predicted (Glassmeier and Espley 2006) and observed by MESSENGER (Sundberg et al. 2012a, Liljeblad et al. 2014). By using MESSENGER data, Liljeblad et al. (2015) showed that the local reconnection rate was very low at the magnetopause crossing associated with the presence of a low-latitude boundary layer (LLBL), ruling out direct entry by local reconnection as a layer formation mechanism. In fact, at Mercury KH waves (which have been suggested to provide particle entry into the LLBL at Earth; e.g. Nakamura et al. 2006) predominantly occur at the dusk side of the magnetopause where, due to kinetic effects resulting from the large gyroradii, ions counter-rotate with the waves (Sundberg et al. 2012a, Liljeblad et al. 2014). On the opposite flank ions co-rotate with the waves resulting in reduced growth rates and larger LLBL (Liljeblad et al. 2015). Mid-latitude reconnection associated with KH instabilities was also discussed by Faganello and Califano (2017), and Fadanelli et al. (2018). In fact, Gershman et al. (2015b) showed that a subset of nightside KH vortices actually has wave frequencies close to the Na$^+$ ion gyrofrequency, indicating that those ions can alter KH dynamics, probably through kinetic effects. James et al. (2016) showed evidence that the KH instability is driving Hermean ULF wave activity that could be better identified by BepiColombo instrumentation (magnetometers and charged-particle detectors). ULF waves are also used to estimate plasma mass density profiles along field lines and, generally, within the magnetosphere (James et al. 2019) complementing observations made by particle instrumentation.

Another peculiarity of the Hermean magnetosphere compared to the terrestrial one is that the plasma density gradients are observed to have much smaller spatial scales and are much more pronounced than that on Earth, due to the interaction of a strongly choked plasma (the solar wind) with a nearly empty cavity constituted by the small-scale magnetosphere.

The role of the instability of the density gradient in the planet-Sun interaction is an interesting topic to investigate also as an example of other comparable exoplanetary environments. BepiColombo will enable analysis of the smaller scales instabilities or "secondary" instabilities that may be much more efficient than fluid-scale instabilities in plasma mixing processes (Henri et al. 2012, 2013). One of those mechanisms could be the lower-hybrid induced, non-adiabatic ion motion across the magnetopause. Since the lower hybrid waves are almost electrostatic, it was not possible to test this hypothesis with MESSENGER. Some coordinated observations between the two BepiColombo spacecraft are suggested in Sect. 4.2.

A further important aspect of solar wind/magnetospheric plasma mixing is the phenomena associated with violent solar wind events. The proximity of Mercury to the Sun compared to the Earth and the small scale of its magnetosphere make it even more responsive to unusually strong events (see next Sect. 2.4).

Another key question concerns whether the mechanism of impulsive penetration observed at Earth (e.g. Echim and Lemaire 2002) is operating at Mercury. We know that at least the small-scale variation in the momentum and density necessary for this mechanism exists in the form of magnetic holes (Karlsson et al. 2016). These findings should be further investigated with linked observations from MPO and Mio, because these will bring a better understanding of the space weathering at Mercury and its contribution to the generation of Mercury's exosphere, as detailed in Sect. 2.7.

## 2.4 How do the Solar Wind and Planetary Ions Gain Energy, Circulate Inside the Magnetosphere and Eventually Impact the Planetary Surface? What Is the Current System in Mercury's Magnetosphere?

As explained in Sect. 2.3, solar wind plasma enters the magnetosphere through dayside magnetopause reconnection, as at Earth, but this process takes place at Mercury even when the magnetic shear angle, the angle between the IMF and planetary magnetic field (Di Braccio et al. 2013) is low. The time resolution (10 s) and the angular coverage ($1.15\pi$) of MESSENGER particle measurements was insufficient to study the resulting acceleration, while BepiColombo will provide resolutions up to 4 s and full angular coverage (see Sect. 4.5). Newly-reconnected magnetic field lines containing solar wind plasma are convected through the magnetospheric cusps to form the plasma mantle, where the competition between downtail motion and E x B drift toward the central plasma sheet determines which solar wind ions end up in the lobes of the magnetotail (Di Braccio et al. 2015b; Jasinski et al. 2017). Solar wind $H^+$ and $He^{2+}$ that make it to the central plasma sheet retain their mass-proportional heating signatures when observed there. Reconnection between lobe magnetic field lines in the tail sends plasma sheet ions and electrons both tail-ward and planet-ward, where they escape downtail, impact the surface or are lost across the magnetopause. Precipitating ions observed on the nightside are mainly at mid- to low- latitudes where the magnetic field lines are closed, that is, with both ends connected to the planet, (Korth et al. 2014), providing evidence of a relatively large loss cone (Winslow et al. 2014). The magnetic flux carried in this process is returned to the dayside completing the Dungey cycle (Dungey 1961) in a few minutes (Slavin et al. 2010, 2019b).

At Earth, energetic charged particles trapped inside the planetary magnetic field azimuthally drift around the planet, because of gradient and curvature drifts. The drift paths are along the iso-contours of the magnetic field. A consequence of the small Hermean magnetosphere with a relatively big portion occupied by the planet is the lack of a significant ring current (Mura et al. 2005; Baumjohann et al. 2010). In fact, closed drift paths around the planet are not allowed.

MESSENGER provided evidence of planetary ions in various regions of the Mercury's magnetosphere (e.g., Zurbuchen et al. 2011; Raines et al. 2013), primarily in the northern magnetospheric cusp and central plasma sheet. Sodium ions originating from the sodium exosphere are one of the main contributors of planetary ions to the magnetospheric plasma at Mercury. Such processes have been explored with statistical trajectory tracing in the electric and magnetic field models of the Mercury's magnetosphere either using empirical (e.g., Delcourt et al. 2007, 2012) or MagnetoHydroDymanic (MHD) simulations (e.g., Seki et al. 2013; Yagi et al. 2010, 2017). Even with steady magnetospheric conditions, the dynamics of sodium ions can change dramatically with conditions of the surface conductivity (Seki et al. 2013) or solar wind parameters (Yagi et al. 2017). Various mechanisms can contribute to the energisation of sodium ions, including $i$) acceleration by convective electric field around the equatorial magnetopause, resulting in the partial sodium ring current (Yagi et al. 2010), $ii$) the centrifugal effect due to curvature of the electric field drift paths (Delcourt et al. 2007), and $iii$) induction electric field during substorms (Delcourt et al. 2012). Some evidence consistent with centrifugal acceleration has been observed, such as sodium ions being predominantly observed in the pre-midnight sector of the magnetotail (Raines et al. 2013; Delcourt 2013) but observations of ions undergoing such acceleration have not been reported. The much more comprehensive instrument complement on BepiColombo, including mass spectrometers on both spacecraft, should enable these concrete connections between models and observations to be established (see Sect. 4.5).

Waves should play a substantial role in particle acceleration at Mercury, due to the highly dynamic nature of Mercury's magnetosphere. MESSENGER observations of wave activity at different frequencies (Boardsen et al. 2009, 2012, 2015; Li et al. 2017; Sundberg et al. 2015; Huang et al. 2020) indicates different physical processes at work and has shown the expected turbulent cascade of energy from MHD scales down to ion-kinetic scales. Much of this power spectrum lies within the ion-kinetic regime, and so wave-particle interactions like ion cyclotron damping should play a role in the acceleration of both solar wind and planetary ions within the system. However, understanding of particle acceleration through such mechanisms remains limited, requiring further treatment by both theory and numerical modelling. Ultimately, BepiColombo measurements will resolve magnetospheric wave activity in considerably more detail and enable significant progress towards answering the question of how both ions and electrons are accelerated (see Sect. 4.5).

While much has been learned about ions >100 eV from MESSENGER observations, low-energy planetary ions are a complete mystery. Born around 1 eV, these ions have never been observed as MESSENGER's lower energy bound was 46 eV for most of the mission (Raines et al. 2014). If present, these ions could have substantial effects. At Earth, such low-energy ions have been shown to alter the kinetic physics of magnetic reconnection on the dayside magnetopause (Borovsky and Denton 2006; Li et al. 2017). Studies of field line resonances show that the total plasma mass density on the dayside may be >200 AMU/cm$^3$, in contrast to the very low (sometimes undetected) densities measured by MESSENGER (James et al. 2019). In the magnetotail, a substantial cold planetary ion population in the central plasma sheet would substantially change the mass density and may be one of the unseen factors causing asymmetries observed there, in reconnection signatures (Sun et al. 2016), current sheet thickness (Poh et al. 2017), and field line curvature (Rong et al. 2018). Thus far, none of these asymmetries have been tied to the 0.1–10 keV planetary ions observed in this region by MESSENGER. BepiColombo/Mio will be able to measure the spacecraft potential, thus estimation of the density of the lower-energy ions could be obtained.

At the Earth, there are two primary large-scale current systems which flow into/out of the high latitude ionosphere, known as region 1 and region 2 currents. These currents specifi-cally couple the magnetopause and the inner magnetosphere, closing through Pedersen currents in the ionosphere (Fig. 6a), and are enhanced during periods of high magnetospheric activity. The region 1 currents map to higher latitudes in the ionosphere and are upward on the dusk side and downward on the dawn side. The region 2 currents map to locations equatorward of this and have opposite polarity. Mercury's magnetosphere differs from that of the Earth for the smaller size relative to the planetary radius, the higher amplitude and smaller timescales for magnetospheric dynamics at Mercury, and the lack of a conducting ionosphere for current closure. Glassmeier (2000) suggested that current closure is not required in any ionosphere and it is possible in the magnetospheric plasma proper, while early simulations predicted that in this small magnetosphere region 1 currents could develop and close inside the highly conductive planetary interior while the region 2 currents could not fully develop (Janhunen and Kallio 2004). Field-aligned currents are typically observed at the Earth using magnetic field measurements taken by spacecraft passing over the high latitude regions. A model of the internal magnetic field is subtracted from each pass, and the residual magnetic field is analysed for perturbations indicative of a local current. Anderson et al. (2014) performed this analysis on Mercury's magnetic field using MESSENGER data and concluded that region 1 field-aligned current signatures were identifiable, particularly during geomagnetically active times (Fig. 6b). These signatures suggested typical total currents of 20–40 kA (up to 200 kA during active times), which may be compared with current strengths ∼MA at the Earth. The signatures were relatively smooth and occurred on every orbit passing through the current regions, which implied that the current systems were stable. This raises an important open question of current stability, given the short timescales for dynamics at Mercury (e.g. Slavin et al. 2012; Imber and Slavin 2017).

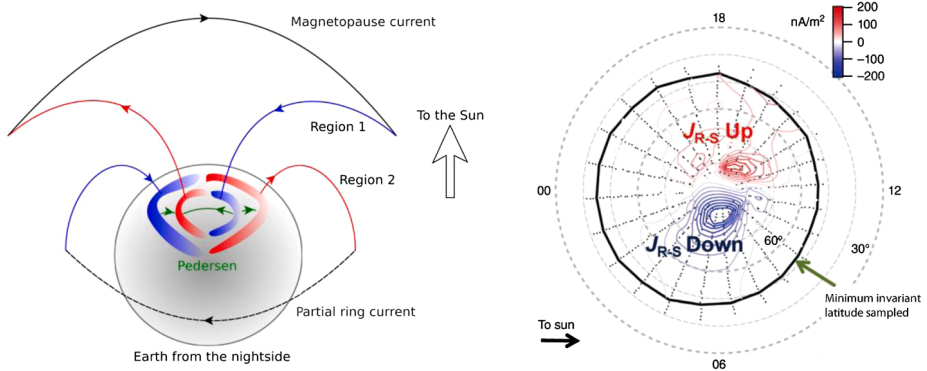

**Fig. 6** (**a**) A schematic of the large-scale field-aligned currents observed in the northern hemisphere at the Earth, from Carter et al. 2016. (**b**) An example of northern hemisphere field-aligned currents accumulated over many MESSENGER orbits during 2012, from Anderson et al. 2018

While Mercury does not have a conducting ionosphere, it does have a large metallic core with a radius of ∼0.8 $R_M$, above which is a lower conductivity silicate mantle. This unique topology allows field-aligned currents to close across the outer surface of the core (see also Sect. 2.2). One of the key open questions for BepiColombo in the realm of magnetospheric dynamics is the closure mechanism for region 1 field-aligned currents and the extent to which induction currents are driven at Mercury.

Finally, Anderson et al. (2014), in agreement with previous modelling results (Janhunen and Kallio 2004), did not find evidence for region 2 currents at Mercury, suggesting that plasma returning to the dayside from the magnetotail may impact the surface. This sug-gestion is strongly supported by observations of MESSENGER/X-Ray spectrometer (XRS) which show evidence of X-ray emission from Mercury's nightside surface, mainly located between 0 and 6 h local time (Lindsay et al. 2016), caused by fluorescence attributed to precipitation of electrons originating in the magnetotail (Starr et al. 2012) (Fig. 7a). Electrons around 10 keV were observed in association with magnetic field dipolarisations in the magnetotail (Dewey et al. 2018). The offset dipole magnetic field at Mercury is expected to cause an asymmetry in the north-south precipitation intensity and location, however MESSENGER's elliptical orbit did not allow all regions of the surface to be equally accurately characterised.

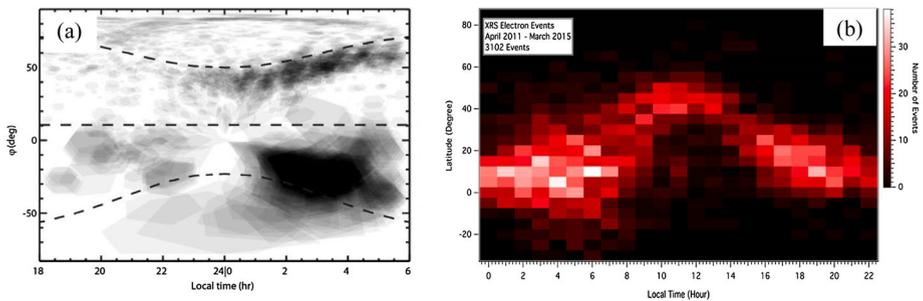

**Fig. 7** (**a**) Maps of XRS footprint locations associated with XRS records containing magnetospheric electron-induced surface fluorescence in latitude-local time coordinates centered at midnight. The southern hemisphere data have a lower spatial resolution. (Lindsay et al. 2016). (**b**) The distribution of suprathermal electron events by latitude and local time centered at noon. The events spanned all local times but had the highest concentration in the dawn and dusk sectors (Ho et al. 2016)

Observations of the loading and unloading of open magnetic flux in Mercury's magnetotail (Imber and Slavin 2017), combined with in situ measurements of reconnection-related phenomena such as dipolarisation fronts (Sundberg et al. 2012a, 2012b), flux ropes (e.g. Smith et al. 2017) and accelerated particles (Dewey et al. 2017), reproduced by ten-moment multifluid model (Dong et al. 2019) (Fig. 4 upper panel), conclusively demonstrate that reconnection signatures may be routinely observed by a spacecraft passing through the magnetotail at a down tail distance of 1–4 $R_M$. Furthermore, many observations of these reconnection-driven phenomena demonstrated a significant dawn-dusk asymmetry, with the majority being observed in the dawn sector, reproduced by MHD-EPIC model (Chen et al. 2019). This is particularly intriguing given that equivalent observations in Earth's magnetotail are offset towards the dusk sector (e.g. Imber et al. 2011). This scenario is compatible with the observed X-ray emissions, but the results thus far are inconclusive. This partly due to the limited nature of the electron observations, which were observed indirectly with MESSENGER's Gamma Ray Spectrometer (Goldsten et al. 2007; Lawrence et al. 2015).

The spatial extent and frequency of precipitation toward the surface, along with the intriguing differences between the predicted and observed hemispheric asymmetries will be targeted by BepiColombo (see Sect. 4.7). The electron in situ observations coupled with X-ray remote observations and the full coverage of the two hemispheres performed by Bepi-Colombo will contribute to determining the conductivity profile of the interior of the planet, investigating the stability of the currents and their response to extreme magnetospheric dynamics, and searching for possible conditions under which region 2 currents may develop.

## 2.5  What Is the Effect of Solar Energetic Particles on Mercury?

MESSENGER studies established an abundance of quasi-trapped energetic electrons in Mercury's magnetosphere. Though MESSENGER did regularly observe energetic electrons in the 35–100 keV range, with excursions to 200 keV (Ho et al. 2012) (Fig. 7b), the expected connections with magnetospheric activity have yet to be established. These electrons are most likely associated with an inductive electric field resulting from the rapid reconfiguration of the magnetic field at reconnecting X-lines (Slavin et al. 2018). Because of the relatively small size of the Hermean magnetosphere, these substorm-injected electrons are often unable to complete a full orbit around the planet in the azimuthal direction before being lost. This means that, in contrast to all other planets with an internal magnetic field, no "Van Allen"-like radiation belts are formed.

During large solar energetic particle (SEP) events (associated either with CMEs or solar flares), a significant portion of the high-energy particles will have direct access to the closed-field-line inner magnetosphere. Ions and electrons populations could form and be maintained for hours after a SEP arrival at Mercury, with a significant dawn/dusk charge separation (Leblanc et al. 2003). This allowed Gershman et al. (2015a) to use 11 SEP events measured by the FIPS instrument to map in detail the extent (and predicted day-night asymmetry) of Mercury's northern polar cap as a function of local time.

The interaction of SEPs (protons, helium and heavier nuclei, abundant mainly in large SEP events—e.g., Desai and Giacalone 2016) with Mercury's surface can create a variety of secondary products, including neutrals, photons and secondary charged particles. Once the secondary products are created at the surface, they contribute to Mercury's exosphere and to magnetospheric plasma mass loading in a direct or indirect way (e.g., ionisation of the generated exosphere).

Composite measurements of particles and radiation at different wavelengths by Bepi-Colombo will offer the opportunity to fully explore the interplay of these particles with the dynamic Hermean environment (see Sect. 4.7). During periods of intense solar activity BepiColombo will provide important information on the space weather conditions around Mercury. Such a feedback is expected to bring a scientific return that goes beyond the scope of a single mission, integrating, for instance, the efforts of other Solar System missions, such as the ESA Solar Orbiter, the NASA MAVEN and Parker Solar Probe missions.

Particle-induced X-ray emission (PIXE) is of especially strong interest for the Bepi-Colombo mission to characterise the structure and composition of Mercury's surface (Huovelin et al. 2010). Proton cross-sections for PIXE peak at proton energies of 1.5–15 MeV for lines of the most interesting elements (Huovelin et al. 2010). The PIXE production for those lines from typical proton spectra (rapidly decreasing at high energies) observed during large SEP events peaks at energies from below 1 MeV to about 5 MeV (Harjunmaa 2004). At these energies, protons are still subject to considerable shielding effects by the Hermean magnetic field. Applying a dipole-field shielding formula (Størmer cut-off) shows that the regions close to the equator are largely inaccessible to particles from outside the magnetosphere: protons with rigidity <55 MV cutoff (equivalent of ≈2 MeV) cannot reach Mercury's surface in the latitude range −20°/+20°; particles with rigidity <20 MV cutoff (≈200 keV) cannot reach it in the latitude band −40°/+40° (Laurenza 2011). According to simulations in a simplified model magnetosphere (Kodikara et al. 2011), the Hermean magnetosphere has a significant effect on precipitating particle trajectories even at energies that are clearly above the cutoffs. Cutoff rigidities computed for different shapes of the Mercury's magnetosphere, responding to different solar wind conditions, through a Toffoletto-Hill modified model (Massetti et al. 2017), showed North/South and dawn/dusk asymmetries in the particle access. Therefore, it is important to combine modelling of the energetic charged particle transmission through the Hermean magnetosphere with multi-spacecraft observations by BepiColombo to fully understand the evolution of the ∼MeV SEP flux spatial distribution inside the Hermean magnetosphere and in the near-surface regions (see Sect. 4.4).

At the highest proton energies, i.e., at some tens to a hundred MeV, protons start to contribute to the gamma-ray production both at the planetary surface as well as within spacecraft structures. Gamma-ray spectroscopy of the planetary surface would typically try to avoid periods with high fluxes of the most energetic SEPs (Peplowski et al. 2012). At such high energies, the field is no longer able to significantly alter the particle trajectories, so that there are no shadowed regions and a measurement of, e.g., >30 MeV proton flux at any point in the Hermean system will give relatively accurate information of the presence of such protons in the whole environment. Thus, estimating the flux of gamma-ray-producing protons in the Hermean system is a somewhat simpler problem from the point of view of characterising the primary particle environment as compared to the more complex dynamics at lower energies.

## 2.6 What Are the Exosphere Composition and Distribution?

Mercury's surface-bounded exosphere is generated by the interaction of the surface with different drivers, such as ions, electrons, meteoroids, photons and thermal radiation. The surface release processes considered as possibly responsible for the exosphere generation at Mercury are photon-stimulated desorption (PSD), ion sputtering, micrometeoroid impact vaporization (MIV), electron stimulated desorption (ESD) or direct thermal release (e.g., Milillo et al. 2005; Killen et al. 2007; Seki et al. 2015). These drivers have different efficiencies depending on the species and how they are bonded with other molecules. Generally, refractories are responsive only to the most energetic processes like MIV and ion sputtering, while volatiles are sensitive to the intense thermal and UV radiation due to Mercury's proximity to the Sun. Most released particles have ballistic orbits and fall back onto the surface (sticking or bouncing again), but some can interact with the solar radiation in different ways after their release. Radiation pressure is effective in accelerating some species, primarily Na and K, in the anti-solar direction, shaping a tail that is modulated by the variation of the radiation pressure along Mercury's orbit (being minimum at perihelion and aphelion, and maximum at the mid seasons) (e.g.: Smyth and Marconi 1995; Baumgardner et al. 2008; Schmidt et al. 2012) (Fig. 8). Other species can be quickly photo-ionised and begin circulating in the magnetosphere as planetary ions. The released atomic groups (mainly after MIV) can be further dissociated, gaining energy. For a full characterisation, multiple instruments and systematic observations of the exosphere at different conditions are required, as well as a combination of simultaneous measurements of possible drivers and the resultant final particles, such as photo-ionised ions. BepiColombo will simultaneously observe the exospheric composition, solar wind, planetary ions and dust (see Sects. 4.6, 4.8, 4.9 and 4.11).

The exosphere of Mercury was discovered by the Mariner 10 Ultraviolet Spectrometer (Broadfoot et al. 1974), which measured H and He and obtained an upper limit for O (Shemansky 1988). Hydrogen was also measured by the Ultraviolet and Visible Spectrometer (UVVS), a subsystem of MASCS, on board of MESSENGER (McClintock et al. 2008; Vervack et al. 2018). Although the scale heights of H measured by Mariner 10 and MESSENGER agree, consistent with a temperature of ∼450 K, the intensities measured by MASCS were a factor of about 3–4 greater than those measured by Mariner 10. The surface number density, $n_0$, of H inferred from the first MESSENGER flyby was $70 < n_0 < 250$ cm$^{-3}$, while that inferred from the second flyby was $65 < n_0 < 95$ cm$^{-3}$.

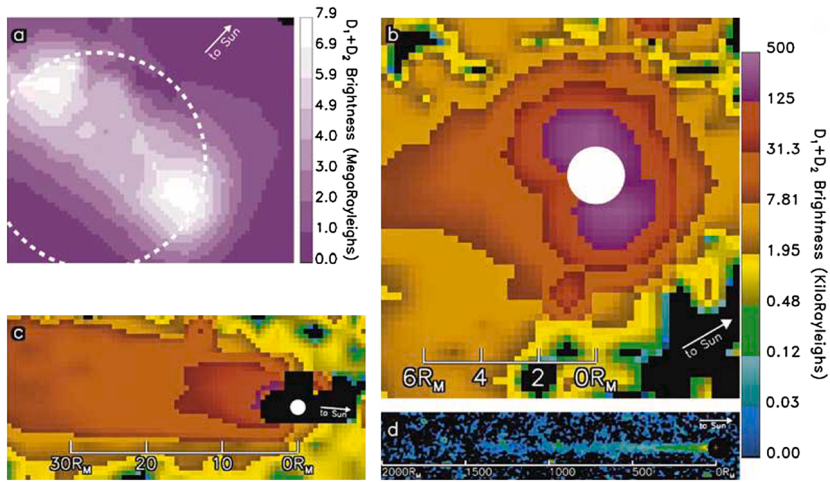

**Fig. 8** A composite of four images of sodium at Mercury showing spatial scales ranging from the diameter of the planet, to approximately 1000 times that size. Image obtained using (**a**) the 3.7 m AEOS telescope on Maui on 8 June, 2006, and (**b**) the 0.4 m telescope at the Tohoku Observatory on Maui on 10 June, 2006. (**c, d**) Obtained using the 0.4 m and 0.1 m telescopes at the Boston University Observing Station at the McDonald Observatory on the night of 30 May, 2007. The tail brightness levels at distances larger than $\sim 10 \ R_M$ are higher in Figs. 1c than 1b, a manifestation of exospheric variability. (Baumgardner et al. 2008)

MESSENGER did not observe He since the wavelength range of MASCS did not extend to the 58.4 nm He emission line. Mariner 10 obtained a maximum He column density of $2.5 \times 10^{12}$ cm$^{-2}$, and a single scale height consistent with $T = 450$ K. One intriguing point of the Mariner 10 observations is the mismatch between the models and the altitude profile closer to the terminator, while altitude profiles closer to the subsolar point were reproduced accurately by the models (Broadfoot et al. 1976). This mismatch was interpreted by Shemansky and Broadfoot (1977) and Smith et al. (1978) to be due to a poorly understood thermal accommodation, which makes the exospheric density more or less dependent on the surface temperature (hence the mismatch closer to the terminator, where surface temperature was less constrained). Some insights have come from the Moon, where helium observations by the orbiters LADEE and LRO can be explained by helium being fully accommodated to the lunar surface (Hurley et al. 2016; Grava et al. 2016). But for Mercury, the lack of measurements has significantly impeded further progress in understanding the gas-surface interaction, a fundamental parameter in the study of temporal evolution of exospheres. The UV and mass spectrometers on board BepiColombo will be able to detect helium, filling this decade-long gap.

Another interesting aspect of $^4$He is that some unknown fraction of it can come from outgassing from the Hermean interior, since $^4$He is the radioactive decay product of $^{232}$Th, $^{235}$U, and $^{238}$U within the crust. Current data cannot constrain that, but again some insights come from the Moon, where $\sim 15\%$ of the exospheric helium is unrelated to the solar wind alpha particle (He$^{++}$) influx, the main source of helium, and presumed to be endogenic (Benna et al. 2015). BepiColombo will measure the exospheric $^4$He and simultaneously the solar wind alpha particle fluxes, hence it will be able to constrain the Hermean endogenic $^4$He source rate (see Sect. 4.6).

Concerning diffusion from the crust of exospheric species, $^{40}$Ar is another radiogenic gas, being the result of radioactive decay of $^{40}$K in the crust that ultimately finds its way to the exosphere through cracks or fissures (Killen 2002). The Mariner 10 UV spectrometer could only place a generous upper limit of subsolar density of $6.6 \times 10^6$ cm$^{-3}$ (Shemansky 1988), based on the sensitivity of the instrument. This density would make $^{40}$Ar one of the most abundant species in the Mercurian exosphere. Interestingly, on the Moon this is indeed the case: $^{40}$Ar and $^4$He are the most abundant exospheric elements identified so far, peaking at a few $10^4$ cm$^{-3}$ (Hoffman et al. 1973). MESSENGER MASCS bandpass did not include the emission line doublet of $^{40}$Ar at 104.8 and 106.7 nm. A measurement by BepiColombo of the column density of $^{40}$Ar (and hence of its source rate) would constrain the abundance of $^{40}$K within the crust. This measurement coupled with the measurement of the ionised component $^{40}$Ar$^+$, providing the loss rate for this element (photo-ionization and electron impact ionization being the major loss processes), with important implications for Mercury formation (see Sect. 4.11).

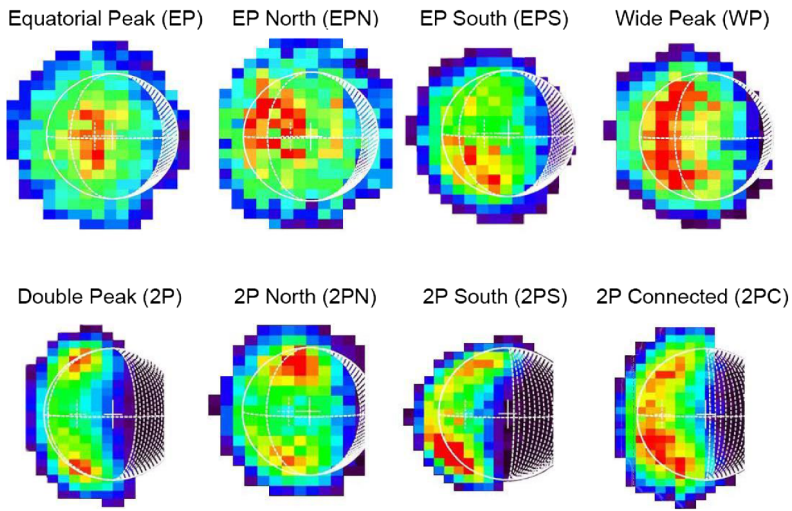

**Fig. 9** Examples of recurrent Na emission patterns identified in the Hermean exosphere. Equatorial Peak South, and Wide Peak, Bottom, from left to right: 2 symmetric peaks, 2 peaks with northern spot dominant, 2 peaks with southern spot dominant and 2 peaks connected. (Mangano et al. 2015)

Although searches for O were regularly conducted by the MASCS instrument on board MESSENGER, there was no clear detection of O in the spectrum. The upper limit for O is ∼2 R (Rayleighs) at 130.4 nm (Vervack et al. 2016). The value reported by Mariner 10 (60–200 R; Broadfoot et al. 1976), well above this number, would have been easily detected during the UVVS observations. It is possible that depletion of oxygen in the Hermean exosphere occurs by condensation of metal oxides and by formation of slowly photolyzed oxides (Berezhnoy 2018) rather than being ejected as neutral atoms. Where is the expected Mercurian oxygen? The identification of atom groups by mass spectrometers on BepiColombo is a unique way for looking for oxides (see Sect. 4.9).

The sodium exosphere of Mercury was first observed from the ground by Potter and Morgan (1985) using the high-resolution echelle spectrograph at the McDonald Observatory. Since that time sodium has been the most observed species in the Hermean exosphere, thanks to its intense intrinsic brightness. North/south asymmetries and variable high latitude enhancements of sodium have been observed with ground-based instruments (e.g. Potter et al. 1999; Killen et al. 1999; Mangano et al. 2009, 2015) (Fig. 9), including transit observations showing northern or southern enhancements at the limb (Schleicher et al. 2004;

Potter et al. 2013; Schmidt et al. 2018). K has been observed by ground-based observations too, showing behaviour similar to that of Na (Potter and Morgan 1986; Killen et al. 2010). An extended sodium tail was first observed by Potter et al. (2002), and subsequently studied by Potter et al. (2007), Baumgardner et al. (2008), Potter and Killen (2008), and Schmidt (2013). These studies demonstrated that the extent of the tail strongly depends on the TAA. Mouawad et al. (2011) showed that the simulated Na exosphere strongly depends on the assumed velocity distributions of the source processes, the composition of the regolith, and sticking and thermal accommodation factors assumed in the simulation. The low to medium-energy source processes such as PSD and MIV are more likely to provide Na to the tail (Schmidt et al. 2012). A fairly repeatable seasonally varying equatorial sodium exosphere was reported by Cassidy et al. (2015). Cassidy also reported a repeatable pattern of East/West sodium asymmetries tied to the Mercury TAA of the planet explained as due to higher Na condensation in the surface regions where the average temperature is colder (cold poles) (Cassidy et al. 2016) (Fig. 10). Leblanc and Johnson (2010) suggested that thermal desorption and photon-stimulated desorption are the dominant source processes for the Na exosphere of Mercury. Recently, Gamborino et al. (2019), analysing the MESSENGER vertical profile of the equatorial subsolar Na exosphere concluded that the main process responsible for Na release in this region seems to be the thermal desorption. In contrast, Orsini et al. (2018), by analysing the Na images obtained by the THEMIS telescope coupled with the magnetic and ion measurements of MESSENGER, reported a variation of Na shape related to an ICME arrival at Mercury, thus linking the ion precipitation to the shaping of the Na distribution (see Sect. 2.4). A multi-process mechanism, involving ion sputtering, chemical sputtering and PSD have been invoked to explain the Na relationship with the precipitation of ions (Mura et al. 2009).

Currently, the scientific community is divided between those favouring an interpretation stating that the two variable peaks are linked to solar wind precipitation and that the Sun's activity is the major driver of the Na exosphere configuration at Mercury (e.g: Killen et al. 2001; Mura et al. 2009; Mangano et al. 2013; Massetti et al. 2017; Orsini et al. 2018), and those favouring the variations being due only to the surface temperatures considered over long time scales or according to position along the orbit (e.g: Leblanc and Johnson 2010; Schmidt et al. 2012; Cassidy et al. 2016) or due to the crossing of the interplanetary dust disk (Kameda et al. 2009). The question is even more tricky considering the question of how this volatile element survived throughout Mercury's evolution history. Multi-point and multi-instrument observations by BepiColombo will provide full characterisation of Na together with possible drivers of its surface release (see Sect. 4.10).

Calcium was discovered in Mercury's exosphere by Bida et al. (2000) using the échelle spectrograph HIRES at the Keck I telescope. It was determined to have a very large-scale height consistent with high temperature (Killen and Hahn 2015); in fact, Killen et al. (2005) suggested that the hot calcium atoms are most likely produced by a non-thermal process. This was verified by the MESSENGER MASCS observations which determined that the calcium is ejected from the dawnside with a vertical density profile that has been interpreted resulting from a characteristic energy of about 6.4 eV, which Burger et al. (2014) converted to a temperature of 70'000 K.

Magnesium was discovered by the MASCS spectrometer on-board the MESSENGER spacecraft during the second flyby (McClintock et al. 2009). The flybys observations, analysed by an exospheric model (Sarantos et al. 2011), are consistent with a source located in the post-dawn equatorial region producing a dual temperature distribution (determined by fitting the vertical profile with a Chamberlain model): hot energetic distribution (up to 20000 K) and cool distribution (less than 5000 K). Retrieved temperatures from the MESSENGER

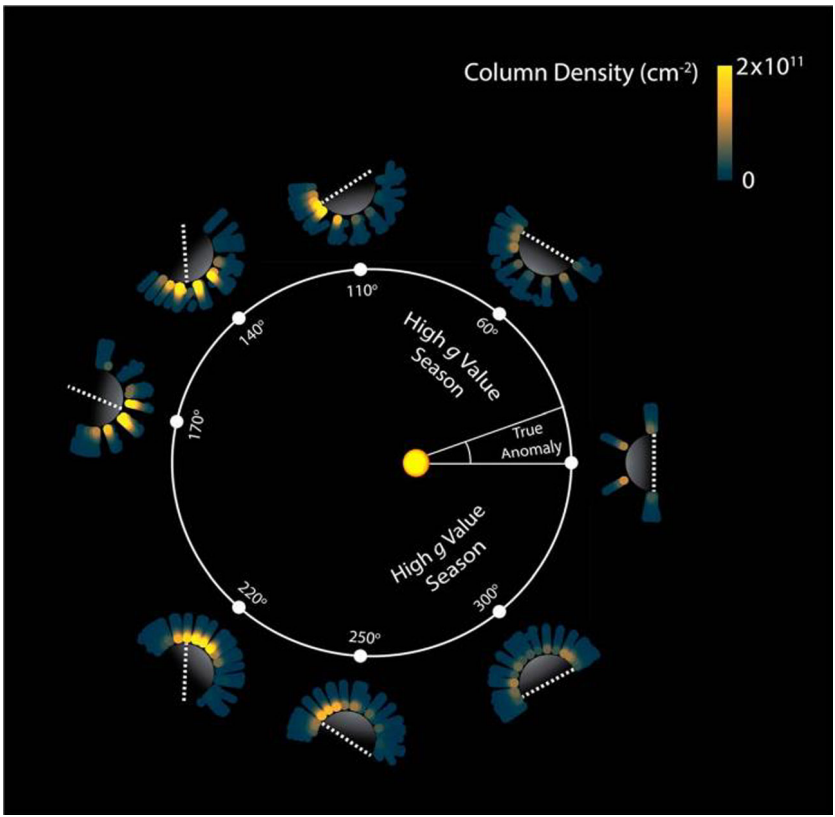

**Fig. 10** Observed sodium column density projected onto Mercury's equatorial plane over the course of one Mercury year. These observations show a sodium enhancement that rotates with the surface and peaks near Mercury's cold-pole longitudes (white dashed lines) when they are sunlit. The enhancement grows over the course of the morning, reaches a peak near noon, and then fades in the afternoon. (Cassidy et al. 2016)

MASCS data along the orbit evidenced periods of a single source in the dayside at betwen 4000–6000 K and a double source near the dawn terminator, as registered during the flyby, for 15% of the time (Merkel et al. 2017).

Observations showed evidence of a dawn enhancement also correlated to the Mg-rich surface region (Merkel et al. 2018). Both Ca and Mg are consistent with impact vaporization in the form of molecules, which are subsequently dissociated by a high-energy process (Killen 2016; Berezhnoy and Klumov 2008; Berezhnoy 2018). The location and timing of the enhanced Ca emission near TAA $= 30°$ are suggestive of a connection with the comet 2P/Encke dust stream (Killen and Hahn 2015; Christou et al. 2015; Plainaki et al. 2017) (Fig. 11), which is also suggested to be the primary driver of Mg, Al, Mn and $Ca^+$ observed at these particular TAA (Vervack et al. 2016).

$Ca^+$ was first detected in Mercury's exosphere during the third MESSENGER flyby (Vervack et al. 2010). Although $Ca^+$ was not regularly detected by UVVS, it was detected on several occasions during the last year of the mission. The FIPS team was unable to unambiguously confirm the detection of $Ca^+$ due to limited mass resolution of the instrument and possible overlap with $K^+$ (Zurbuchen et al. 2008). From modelling the ion measurements of FIPS, it was found that the $Ca^+$ abundance is about two decades larger than the $K^+$ (Wurz et al. 2019).

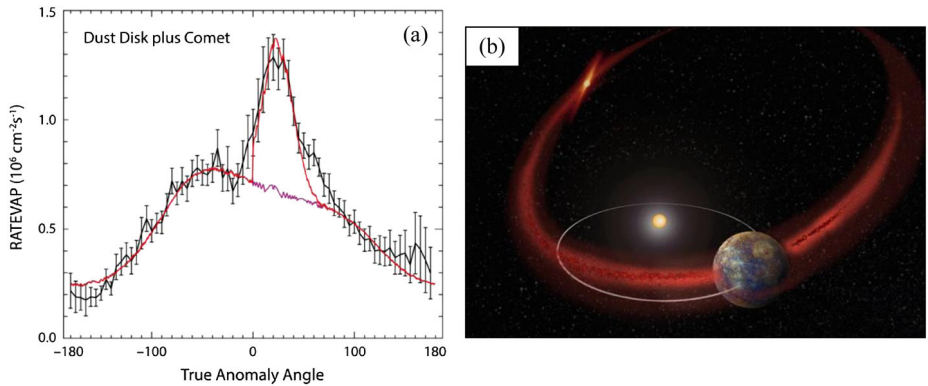

**Fig. 11** (**a**) Ca vaporization rate at Mercury due to the interplanetary dust-disk (magenta line) plus a cometary stream whose peak density occurs at TAA 25°. The red line is the summed contributions from the cometary dust stream plus that due to an interplanetary dust-disk that is inclined 10° from Mercury's orbital plane, and whose ascending node is 290° when measured from Mercury's longitude of perihelion, with the dust density varying as $R^2$, where $R$ is the heliocentric distance. The MASCS observations are plotted in black. (**b**) Sketch of the Mercury orbit crossing the 2P/Encke dust stream. (Killen and Hahn 2015)

Bida and Killen (2011) reported measurements of Al at line-of-sight abundances of $(2.5–5.1) \times 10^7$ cm$^{-2}$ from 860 to 2100 km altitude from observing runs at the Keck 1 telescope during 2008 and 2011 (Bida and Killen 2017). Al was also detected by MASCS late in the MESSENGER mission, at a line-of-sight column abundance of $7.7 \times 10^7$ cm$^{-2}$. The UVVS value pertains to lower altitudes (250–650 km) than those measured by Keck and are thus considered consistent.

The UVVS data revealed the unexpected presence of Mn at an estimated column of $4.9 \times 10^7$ cm$^{-2}$ (Vervack et al. 2016), but it is estimated to be highly variable. Because the geometry of the observation was complicated, the column abundance is considered as an order-of-magnitude.

In conclusion, Mariner-10, MESSENGER, and the numerous ground-based observations proved the presence of H, He, Na, K, Ca, Mg, Al, Mn in Mercury's exosphere, but other species and atom groups are expected to be present. The BepiColombo multi-type of instrument approach to the exosphere identification (combining remote sensing and in-situ measurement) warrants that new elements will be added to the list (see Sect. 4.11).

## 2.7 What Are the Relationships Between the Solar Wind or the Planetary Ions and the Exosphere?

As anticipated in the previous sections, Mercury's weak magnetic field, the high reconnection rate, and its exosphere allow the solar wind to reach a large portion of the dayside surface, primarily focusing at the base of the cusp regions. However, as solar wind is conflated inside the magnetosphere, it can impact the surface in other regions as well, such as the dawn polar regions (Raines et al. 2014, 2015). The planetary ions directly released from the surface or resulting from the exosphere photo-ionization, circulate and are accelerated in the magnetosphere, and can be convected back onto the surface mainly on the night side at middle latitudes, but also in the dusk flanks and the dayside (Raines et al. 2013, 2015; Wurz et al. 2019) (see Sect. 2.4). Therefore, these charged particles of solar wind or of planetary origin can impact the surface over a wide range of local times. The impact of an energetic ion onto a surface can have a number of different, inter-related consequences, including: the reflection of the neutralized impacting particle, the ejection of neutrals and charged particles from the surface (ion-sputtering process), the production of X-rays in the case of higher energy and high charge-state ions as in the SEP events, and the alteration of the chemical properties of the surface, causing the so-called "space weathering" (Domingue et al. 2014; Strazzulla and Brunetto 2017; see also Rothery et al. 2020, this issue). The ion-sputtered neutral particles contribute to filling the exosphere, but the contribution of this process with respect to other surface-release processes is still unclear and remains a matter of debate within the science community.

While MESSENGER observed ions directed toward the surface and was able to provide estimates of precipitation rates in the northern magnetospheric cusp by analysing deep magnetic field depressions (Poh et al. 2017) and average pitch angle distributions (Winslow et al. 2014). Despite an estimation of the proton precipitation flux in the range $10^6$ $10^7$ cm$^{-2}$ s$^{-1}$ (Raines private communication), MESSENGER was not able to provide a direct proof of the impact onto the surface or the connection with the exosphere generation. The ENA instruments on board BepiColombo will be able to detect the back-scattered ions, thus providing evidence of the impacts (Milillo et al. 2011).

In spite of the rapid changes in the precipitating proton flux due to the fast magnetospheric activity (time scales of 10s of seconds) and magnetic reconnection processes, the exosphere would show a much smoother response, because of the time delay of the exosphere transport. In fact, ballistic time scale is about 10 minutes after MIV (Mangano et al. 2007) while the exosphere requires some hours to recover after a major impulsive ion precipitation event (Mangano et al. 2013; Mura 2012). However, the predicted close connections linking the ion precipitation with clear changes in the exosphere have been elusive. In fact, ground-based observations by THEMIS telescope suggest that the Na exospheric double peaks can show a variability on a time scale smaller than 1-hour, and possibly shorter-term fluctuations of about 10 minutes (Massetti et al. 2017). On the other hand, MESSENGER observations show an equatorial Na exospheric density almost repeatable from year to year, even if a time variability of some 10s of % not related to orbit or planetographic position, but probably linked to transient phenomena, is clearly registered (Cassidy et al. 2015). The analysis of THEMIS observations recorded during the transit of an ICME at Mercury, registered by MESSENGER (Winslow et al. 2015; Slavin et al. 2014) shows that this event could be put in relation with a variation of the global shape of the Na exosphere, i.e.: at ICME arrival time the two peaks seem to extend toward the equator becoming an exosphere almost uniformly distributed throughout the whole dayside (Orsini et al. 2018). This observation suggests that the magnetospheric compression, could notably affect the whole Na exospheric emission by modifying the cusps extension and driving ion impacts at low latitudes. This agrees with the thick and low-$\beta$ plasma depletion layers observed by MESSENGER during extreme conditions (Slavin et al. 2014; Zhong et al. 2015b). Mercury's magnetopause reaches the planet's surface ~30% of the time during ICMEs (Winslow et al. 2017), but even if the magnetopause was not compressed to the planet surface, the ICME, being rich of heavy ions (Galvin 1997; Richardson and Cane 2004) at large gyroradii, could allow the access of ions to the closed-field line regions (Kallio et al. 2008). Since the heavy-ion sputtering yield is higher than the proton one (Johnson and Baragiola 1991; Milillo et al. 2011), a significant enhancement of surface release could be seen during ICMEs or SEPs (Killen et al. 2012).

Even if the influence of plasma precipitation seems the only way to explain the double peak shape and variability of the Na and K exospheres observed from ground-based telescopes (e.g.: Mangano et al. 2009, 2013, 2015, Massetti et al. 2017, Potter et al. 2006), the ultimate mechanisms responsible for such surface release is not unambiguously identified. In fact, the yields (number of particles released after the impact of a single ion) measured in laboratory simulations in the case of few-keV protons onto a rocky regolith surface are too low to explain the observed Na exosphere (Johnson and Baragiola 1991; Seki et al. 2015). A multi-process action was suggested by Mura et al. (2009) for Mercury's exosphere generation and by Sarantos et al. (2008) in the Moon's case, that includes the alteration of the surface properties induced by the impact so that the subsequent action of photons (PSD) is more efficient, but up to now there is no unambiguous evidence supporting it. BepiColombo, having a full suite of particle detectors for the close-to-surface environment and a Na global imager, will allow a unique comprehensive set of measurements of the exosphere and of the ion precipitation in dayside as well as in the nightside (see Sects. 4.6, 4.8 and 4.10).

The nightside surface is also subject to energetic electron precipitation, as suggested by the MESSENGER X-ray observations (Starr et al. 2012). Possible consequences of these impacts, besides the X-ray emission, would be the release of volatile material due to ESD. The observation of the nightside exosphere is particularly difficult by remote sensing UV spectrometers since there is no solar radiation able to excite the exospheric atoms. Bepi-Colombo's X-ray imager, together with its mass spectrometer, would provide a new important set of measurements (see Sect. 4.8).

The magnetospheric ions of solar wind origin circulating close to the surface may experience another possible interaction with the exosphere that has never been investigated through observations in Mercury's environment: where the exospheric density is higher, an ion can charge exchange with a local neutral atom. The product of the interaction is a low-energy ion (the previously neutral atom) and an ENA (the neutralised energetic ion) having almost the same energy and direction of the parent ion (Hasted 1964; Stebbings et al. 1964). The collection of the charge-exchange ENA generated along a line-of-sight will provide a threefold information: a) remote sensing of the plasma population circulating close to the planet; b) the signature of exospheric loss; and c) a source of planetary ions (Orsini and Milillo 1999; Mura et al. 2005, 2006). BepiColombo, having two ENA sensors, will provide observations of ENA from different vantage points, allowing a kind of reconstruction of the 3D ENA distribution (see Sects. 4.5 and 4.11).

## 2.8 What Effects Does Micrometeoroid Bombardment onto the Surface Have on the Exosphere?

The MIV process results in release of solid, melt and vapor from a volume where the meteoroid hits the surface (Cintala 1992). The released vapour leaves the surface with a Maxwellian energy distribution (corresponding to temperature between 1500 and 5000 K), thus the exosphere is refilled with a cloud contributed by the surface material depending on the energy of vaporization, and gravitationally differentiated (Berezhnoy 2018). Models of surface-bounded exospheres have extremely varied estimates of the importance of impact vaporization, and also vary by degree of volatility of the species. Although impact vaporization is an established field of study (Melosh 1989; Pierazzo et al. 2008; Hermalyn and Schultz 2010) uncertainties regarding the importance of impact vaporization on extra-terrestrial bodies include the uncertainty in impact rates for both interplanetary dust and larger meteoroids and comets (Borin et al. 2010, 2016, Pokorný et al. 2018), the relative amounts of melt and vapour produced in an impact (Pierazzo et al. 1995), the temperature of the vapour—which affects escape rates (e.g. Cintala 1992; Rivkin and Pierazzo, 2005), the relative amount of neutral versus ionized ejecta (Hornung et al. 2000), and the gas-surface interaction of the downwelling ejecta (Yakshinskiy and Madey 2005). Finally, the enhanced volatile content of the Mercurian regolith measured by the MESSENGER instruments requires a reanalysis of previous models.

At Mercury, observations of Ca and Mg exospheres seem to indicate clearly that MIV is the primary responsible process of their generation (see Sect. 2.6). In fact, the identified source is located in the dawn hemisphere where higher micrometeoroid impact flux is expected. The Ca column density increases where the 2P/Encke comet meteoroid stream is expected to cross the Mercury's orbit (Killen and Hahn 2015). The importance of impact vaporization as a source of exospheric neutrals has been constrained in part by observation of the escaping component of the exospheres—the Mercurian tail (Schmidt et al. 2012), the Ca exosphere of Mercury (Burger et al. 2014) and the lunar extended exosphere and tail (Wilson et al. 1999; Colaprete et al. 2016). These results all depend critically on the assumed temperature or velocity distribution of the initial vapour plume, on the assumed photoionization rate and on the interaction of the material with radiation (which is species-dependent). In fact, after release, the material can be subjected to other processes like dissociation or photoionization. Estimates of the Na photoionization rate have varied by a factor of three, and values of the Ca photoionization rate have varied by a factor of about four (e.g. see Killen et al. 2018). This obviously introduces a huge uncertainty in the escape rate, and hence the source process. The quenching temperature of the cloud defines the final constituents of the vapour cloud (Berezhnoy and Klumov 2008; Berezhnoy 2018). The hypothesis of energetic dissociation of atom groups (Killen 2016) considered for explaining the high temperature of the two refractories, Ca and Mg, observed at Mercury exosphere, is unable to fully explain the observed intensity mainly due to uncertainty in the physics of dissociation processes (Christou et al. 2015; Plainaki et al. 2017).

The evaluation of the residence time of the material in the exosphere is an open issue. Colaprete et al. (2016) conclude, based on observations of the UV spectrometer onboard the LADEE mission to the Moon, that released MIV material persists in the exosphere-surface system for much longer than the ionization lifetime. Residence times in the lunar environment of 45 to 90 days (mainly on the lunar surface) can be expected before escape to the solar wind, which would explain the long-term smooth increase and decrease in the Na column density observed as the result of meteoroid streams. This is the result of each particle residing in the regolith for approximately an ionization lifetime (i.e., several days) between bounces, combined with the many bounces that it has to take before being lost from the exosphere. Given the long residence time of Na on the surface deduced by both Leblanc et al. (2003) and Colaprete et al. (2016), it has been suggested that ion or micrometeoroid impacts are the primary source of atoms migrating from the regolith to the extreme surface, and these atoms feed the subsequent release by photons or thermal processes (Mura et al. 2009; Killen et al. 2018). If this is the case, then micrometeoroid impacts play the dominant role in maintaining the exospheres while the less energetic processes such as PSD and thermal desorption serve to keep the atoms in play until they are destroyed by photoionization.

Schmidt et al. (2012) studied the extended sodium tail of Mercury and concluded that both photon-stimulated desorption and micrometeoroid impacts are required to simulate the ~20% loss of Mercury's sodium atmosphere, depending on orbital phase, and that the two mechanisms are jointly responsible for the observed comet-like tail as driven by solar radiation pressure. Furthermore, the Na escape rate derived by the observation of the Moon's distant tail increased by a factor of 2 to 3 during the most intense period of the 1998 Leonid meteor shower (Wilson et al. 1999), evidencing a strong influence of meteor impacts on the lunar sodium exosphere and its escape rate. It was found theoretically that NaO and KO photolysis lifetimes are significantly shorter than its ballistic flight times (Valiev et al. 2020). Planned BepiColombo studies of Na and K content as a function of altitude will be able to estimate properties of photolysis-generated Na and K atoms in the Hermean exosphere (see Sect. 4.9). BepiColombo will offer an unprecedented opportunity to observe simultaneously the exosphere in refractories and volatile components, the Na tail, and the dust environment around Mercury.

Apart from the average exospheric condition, the possibility to register a bigger (1, 10 cm and 1 m) meteoroid impact during the BepiColombo mission has been investigated by Mangano et al. (2007). They concluded that the noticeable increase of some species over the average exospheric density, the amplitude of the resulting cloud, the duration, and the favourable detection probability of the MIV event all work in the direction of positive detections of a cloud from a 10-cm vaporized meteoroid. The vaporized surface hemispherical volume could reach a dimension of meters, depending also on the density and porosity of the regolith; thereby, allowing a kind of remote sensing of the planet endogenous material.

## 2.9 How do the Surface Composition, Mineralogy and Physical Condition Affect the Surface Release Processes?

Surface mineralogy, composition and grain size determine the thermal capacity and the surface release efficiency. MESSENGER observations revealed many structures specific to Mercury. BepiColombo observations will investigate correlations between the surface composition and any discernible patterns in the ratios of species released from the surface thus, having an effect on the exosphere (see also Rothery et al. 2020, this issue).

A new target, not anticipated until they were discovered by MESSENGER, is hollows (Blewett et al. 2018 and references therein; Lucchetti et al. 2018). These are steep-sided, flat-bottomed depressions where the upper 10–20 m of surface material has been somehow lost. They are clearly young, and probably still active today losing volatile components of the surface material at their edge (Blewett et al. 2018). Any enhancement in exospheric species that can be traced back to fields of hollows would help to illuminate the nature and rate of volatile loss. Other possible sites of enhanced volatile loss that might still be ongoing and which should be checked by BepiColombo include conical ejecta blocks surrounding the Caloris basin (Wright et al. 2019), down-slope streaks on the inner walls of a few impact craters and the Nathair Facula vent (Malliband et al. 2019). High spatial and mass resolution measurements of the exosphere and planetary low energy ions above these regions will provide information on the volatile release and on possible changes still active.

Radar-bright units inside permanently-shadowed polar craters have been identified by ground-based radar observations (Harmon and Slade 1992; Harmon 2007; Harmon et al. 2011; Chabot et al. 2018). They were consistent with water-ice or sulfur or supercooled silicates, the MESSENGER neutron spectrometer observations in the northern regions identified hydrogen-rich substance in favor of a water ice interpretation (Lawrence et al. 2013). Reflectance measurements performed by the MESSENGER laser altimeter in the same regions showed surfaces with albedos distinctly different from that surrounding terrain. A few locations have very high albedo consistent with water-ice, but numerous locations have very low albedo, interpreted to be complex carbon-bearing organic compounds (Chabot et al. 2016; Rothery et al. 2020, this issue). In these regions, the surface properties could be quite different from the rest of the planet. In particular, the backscattering rate of solar wind could be higher thus the space weathering effect could be enhanced. The release efficiency could be higher than on the rocky surface (e.g.: Seki et al.

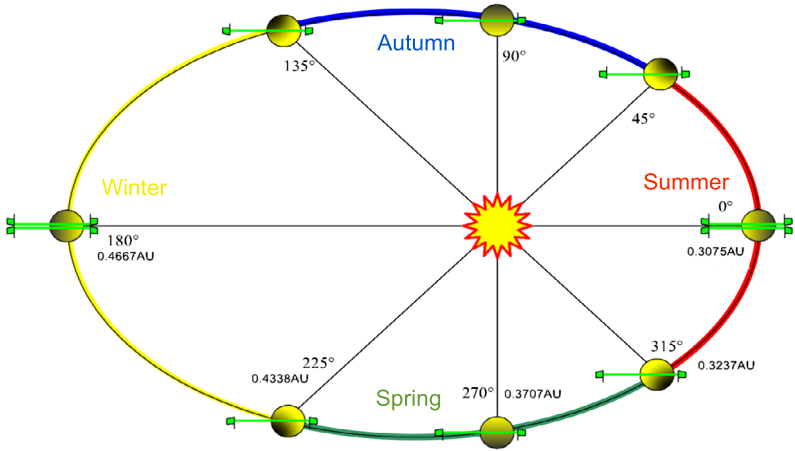

**Fig. 12** BepiColombo orbit phases

2015) and presence of organic material synthetized by GCR (e.g., Lawrence et al. 2013; Paige et al. 2013; see Sect. 2.1) could affect also the yield of the release processes. Finally, an effect on the polar exosphere could be expected, in particular the thermal H component is probably higher at the poles.

MESSENGER could not compare the surface features to the exosphere composition measurements since the MASCS observations did not cover the whole planet at high resolution. Nevertheless, the MASCS data analysis proved a connection between the Mg-rich region (Weider et al. 2015) and the observed local exosphere (Merkel et al. 2018). Eventually, the characterisation of the relationship between the surface properties and the exosphere is a task for BepiColombo, which will provide mapping of visual, NIR, IR, X and gamma spectra (see Sect. 4.12 and Rothery et al. 2020, this issue).

## 3 BepiColombo: An Optimal Mission for Environment Investigation

### 3.1 Mission Configuration in Orbit Phase

#### 3.1.1 Trajectories

The BepiColombo mission (Benkhoff et al. 2020, this issue), launched on 20th October 2018 from the European spaceport in Kourou (French Guiana), is delivering the two spacecraft to Mercury. After the nine gravity assists at the Earth, Venus and Mercury, the MPO and Mio will be separated and will be inserted in the nominal orbits around Mercury between December 2025 and March 2026. MPO will be placed in a polar orbit at 480 km × 1500 km altitude with a 2.3 h orbit period with its apoapsis on the dayside when Mercury is at perihelion, while Mio will be placed in a highly eccentric polar orbit at 590 km × 11 640 km altitude (about 5.8 $R_M$ planetocentric distance) with a 9.3 h orbit period, co-planar with the MPO orbit (Fig. 12).

### 3.1.2 General Science Operations

Even if some scientific operations will be conducted during the cruise phase (Mangato et al. 2020, this issue), BepiColombo nominal BepiColombo scientific operations will start in April 2026. The operation planning will follow different philosophies for each spacecraft. While MPO will consider the instrument requests case by case, with a long-, medium- and a short-term operation plan definition, most of the Mio payload, being entirely dedicated to the fields and particles environment, will use solar wind and and magnetosphere operational modes (Murakami et al. 2020, this issue). Howerver, the MSASI imager (for exospheric sodium) will instead collect on-board about 8 images per orbit, but only one third of them will be downloaded.

The data collected on-board will be downlinked with different time delays for each spacecraft (Montagnon et al. 2020, this issue). For some instruments, a Selective Data Downlink (SDD) approach will be applied, so that the higher resolution data or specific observations collected on board can be requested to be returned via Ka-band for specific observations periods according to the analysis of low resolution data transmitted via the faster X-band. Other instruments will follow a Flexible Data Downlink (FDD) philosophy, whereby they plan in advance periods with part of their telemetry, nominally sent via Ka-band, to be transmitted in the fast X-band. These complexities and differences in data management and downloading require special care and optimal coordination in operation planning, taking into account high-rate data periods and latency (time between generation on board and arrival on ground). A wide view of the scientific goals is crucial for this scope.

### 3.2 Scientific Performances of the Environment Payload

The instruments fully or partially devoted to the study of the Hermean environment that would take advantage of simultaneous or contiguous observations are the whole Mio payload, namely, MGF (Baumjohann et al. 2020, this issue), MPPE (Saito 2020, this issue), PWI (Kasaba et al. 2020, this issue), MDM (Kobayashi et al. 2020, this issue), and MSASI (Yoshikawa et al. 2020, this issue) and, on board MPO, MAG (Heyner et al. 2020, this issue), SERENA (Orsini et al. 2020, this issue), PHEBUS (Quémerais et al. 2020, this issue), SIXS (Huovelin et al. 2020, this issue), MIXS (Bunce et al. 2020, this issue) and a radiation monitor in support of the mission BERM (Moissl et al. 2020, this issue). Table 3 presents a schematic summary of the performances of these instruments.

### 3.3 Inter-Calibration of Instruments That Have Targets in Common

The two spacecraft have some instruments that have the same targets, but different field of view, geometrical factors and efficiencies. Therefore, intercalibration is crucial for the coordinated science objectives that are discussed in Sect. 4. In view of detector aging that will change response functions, intercalibrations should be performed periodically throughout the mission.

Both magnetometers have the same design, but periodic intercalibrations are required, in addition to cross-calibration with the comparative magnetic field measurement made by Mio/PWI.

The low-energy ion sensors of Mio/MPPE, (i.e. MIA and MSA) should be periodically intercalibrated with the MPO/SERENA ion sensors MIPA and PICAM. While the high-energy ions are the target of both Mio/MPPE-HEP and MPO/SIXS-P and the low energy part of BERM.

**Table 3** Main characteristic and performances of BepiColombo payload devoted to the Hermean environment

| Instrument | Target | Main characteristics | Main Scient. Obj. for environment | PIs and Co-PIs |
|---|---|---|---|---|
| *MPO/MAG: 2 Tri-axial fluxgate sensors* | Vector Magnetic Field | res. 2 pT time res. up to 128 Hz | Planetary Magnetic field | *PI*: D. Heyner, Braunschweig University, (Germany) *Co-PI*: C.M. Carr, Imperial College London (UK) *PI emeritus*: K.-H. Glassmeier, Braunschweig University, (Germany) |
| *Mio/MGF: 2 Tri-axial fluxgate sensors* | Vector Magnetic Field | time res. up to 128 Hz res. 4 pT | Magnetic field in the IMF and in the magnetosphere | *PI*: W. Baumjohann, IWF (Austria) *Co-PI*: A. Matsuoka (Japan) |
| *Mio/PWI: Plasma Wave experiment* | *EWO* receiver | Low and Medium frequency electric field; Medium frequency Magnetic field | DC ~ 32 Hz E-field; 10 Hz ~ 120 kHz E-field; 0.1 Hz ~ 20 kHz B-field | Electromagnetic fields, plasma waves, radio waves, electron density and temperature in the Hermean magnetosphere for: 1. Structure of the magnetosphere; 2. Dynamics of the magnetosphere; 3. Energy transfer and scale coupling; 4. Wave-particle interactions; 5. Solar radio emissions and diagnostics. | *PI*: Y. Kasaba, Tohoku University (Japan); *Co-PIs*: H. Kojima, Kyoto University (Japan); S. Yagitani, Kanazawa University (Japan); M. Moncuquet, Observatoire de Paris (France) J.-E. Wahlund, IRF (Sweden) |
| | *SORBET* (Spectroscopic Ondes Radio & Bruit Electrostatique Thermique) receiver | High frequency electric and magnetic field | 2.5 kHz ~ 10 MHz E-field; 10 ~ 640 kHz B-field | |
| | *AM2P* (Active Measurement of Mercury's Plasma) receiver | Antenna impedance measurement and calibration signal source with electron density & temperature measurement | 0.7 ~ 144 kHz (signal output) | |

**Table 3** (Continued)

| Instrument | Target | Main characteristics | Main Scient. Obj. for environment | PIs and Co-PIs |
|---|---|---|---|---|
| *MEFISTO* (Mercury Electric Field In-Situ Tool) and *WPT* (Wire-Probe anTenna) | Electric field | DC $\sim$ 3 MHz E-field; DC $\sim$ 10 MHz E-field | | |
| *LF-SC*: Low frequency search coil | Magnetic field | 0.1 Hz $\sim$ 20 kHz B-field | | |
| *DB-SC* (Dual band search coil) | Magnetic field | 0.1 Hz $\sim$ 20 kHz (L) B-field; 10 $\sim$ 640 kHz (H) B-field | | |
| *MPO/SERENA (Search for Exosphere Refilling and Emitted Neutral Abundances)* | | | 1. Chemical and elemental composition of the exosphere  2. Neutral gas density asymmetries  3. planetary ions composition  4. planetary ions spatial and energy distribution  5. Plasma precipitation and SW distribution in the inner magnetosphere  6. Surface emission rate and release processes  7. Particle loss rate from Mercury's environment | *PI:* S. Orsini, INAF/IAPS (Italy);  *Co-PIs:* S. Barabash, IRF (Sweden); H. Lichtenegger, IWF (Austria); Livi, SWRI (USA) |
| *ELENA* (Emitted Low Energy Neutral Atoms) | ENA: Mapping of the surface back-scattered particles and charge—exchange ENA | Energy range: 20–5000 eV ang. res.: 2° × 2° | | |
| *MIPA* (Miniature Ion Precipitation Analyser) | Solar wind close to the planet | Energy range: 15 eV $-$ 15 keV; Energy and angular discrimination (mode dependent); FOV: 80° × 360°; Rough mass res. | | |
| *PICAM* (Planetary Ion CAMera) | Planetary ions close to the planet | Energy range: 10 eV–3 keV; Energy and angular discrimination (mode dependent); FOV: 1.5$\pi$; Mass res.: $M/\Delta M > 50$ | | |
| *STROFIO* (STart from a ROtating FIeld mass spectrOmeter) | Exosphere | Energy range: 0.01–50 eV; FoV: 20° × 20°; Mass resolution: $M/\Delta M = 60$ | | |

**Table 3** (*Continued*)

| Instrument | Target | Main characteristics | Main Scient. Obj. for environment | PIs and Co-PIs |
|---|---|---|---|---|
| *Mio/MPPE (Mercury Plasma Particle Experiment)* | | | | *PI*: Y. Saito, JAXA/ISAS (Japan); *Co-PIs*: M. Hirahara, Nagoya University (Japan); S. Barabash, IRF (Sweden); D. Delcourt, CNRS – Université d'Orleans (France) |
| *MSA (Mass Spectrum Analyzer)* | Solar wind and planetary ions | Energy range: 1 eV/q–38 keV/q Energy and angular discrimination $4\pi$ coverage Mass res.: $M/\Delta M > 40$ ($<13$ keV/q) $M/\Delta M = 10$ ($>13$ keV/q) | 1. Structure, dynamics, and physical processes (transport, acceleration) in the Mercury magnetosphere | |
| *MIA (Mercury Ion Analyser)* | Solar wind and magnetospheric ions | Energy range: 15 eV/q–29 keV/q Energy spectra $5.625° \times 5.625°$ (Solar wind) $22.5° \times 22.5°$ (Mercury ion) $4\pi$ coverage | 2. Magnetospheric source and loss processes; role and efficiency of the solar wind and planetary surface as sources of plasma for the Hermean magnetosphere 3. Structure and topology of the interplanetary magnetic field lines | |
| *MEA1 and MEA2 (Mercury Electron Analyser)* | Solar wind and magnetospheric electrons | Energy range: 3 eV–25,500 eV (Mercury mode) 3 eV–3000 eV (solar wind mode) Energy spectra $4\pi$ coverage $22.5° \times 11.25°$ | 4. Collisionless shock physics in the inner heliosphere; monitor the solar wind and study interstellar pick-up ions | |
| *HEP-ele and HEP-ion (High Energy Particles)* | High energy electrons and ions | Energy range: 30–700 keV (electrons) Energy range: 30–1500 keV (ions) Energy and angular discrimination Rough mass res. | 5. Investigation of the high energy particles bursts in the magnetosphere | |

**Table 3** (Continued)

| Instrument | Target | Main characteristics | Main Scient. Obj. for environment | PIs and Co-PIs |
|---|---|---|---|---|
| *ENA* (Energetic Neutral Particles) | ENA: back-scattered and charge—exchange ENA | Energy range: 20–5000 eV Energy and angular discrimination | 6. Solar wind precipitation onto the surface and exosphere–magnetosphere interactions | |
| *MPO/SIXS* (Solar Intensity X-ray and particle Spectrometer) *SIXS-X* | Sun X-ray | Spectral range: ~1 keV–20 keV | Monitor the solar X-ray corona and solar flares and to determine their temporal variability and spectral classification | *PI: J. Huovelin,* University of Helsinki (Finland) *Co-PIs: M. Grande,* Aberystwyth University (UK) |
| *SIXS-P* | High energy electrons and ions | Energy range: ~100 keV–3 MeV (electrons) Energy range: ~1–30 MeV (protons) Energy and angular discrimination | Monitoring the solar energetic particle fluxes towards the planet's surface. Investigation of the high-energy particles in the magnetosphere | R. Vainio University of Turku (Finland) |
| *MPO/BERM: Resource spectrometer* | High energy electrons and ions | Energy range: ~0.3–10 MeV (electrons) Energy range: ~1–200 MeV (protons) Energy range: ~1–50 MeV (heavy ions) Energy discrimination FoV: 40° | Support other instruments providing the radiation environment | *PI: R. Moissl, ESA* |

**Table 3** (Continued)

| Instrument | | Target | Main characteristics | Main Scient. Obj. for environment | PIs and Co-PIs |
|---|---|---|---|---|---|
| *MPO/PHEBUS (Probing of Hermean Exosphere by Ultraviolet Spectroscopy)* | EUV FUV NUV | Exospheric emission | Spectral range: 50 nm and 320 nm Spectral res.: between 1 and 1.5 nm NUV channels at 402 nm and 422 nm | Exospheric composition, 3D structure and dynamic Characterisation of the exospheric sources and sinks | *PI*: E. Quémerais, LATMOS-IPSL (France); *Co-PIs*: I. Yoshikawa, University of Tokyo (Japan) O. Korablev, IKI (Russia) |
| *Mio/MSASI (Mercury Sodium Atmospheric Spectral Imager)* | | Na D2 line | Spectral range: $589.158 \pm 0.028$ nm Spatial res.: $0.18° \times 0.18°$ | Abundance, distribution, and dynamics of Sodium exosphere | *PI*: I. Yoshikawa, The University of Tokyo (Japan); *Co-PI*: O. Korablev, IKI (Russia) |
| *Mio/MDM (Mercury Dust Monitor)* | | Impact momentum and direction of dust particles | FoV: $2\pi$ | Study the distribution of interplanetary and ambient dust at the Mercury's orbit. Micrometeoroid impact and surface vaporization Dust sciences of the inner solar system | *PI*: M. Kobayashi, Chiba Institute of Technology (Japan) |
| *MPO/MIXS (Mercury Imaging X-ray Spectrometer)* | MIXS-C MIXS-T | Surface X-ray fluorescence | Spectral range: 0.5–7.5 keV Spectral res.: 140 eV at Fe-K FoV: 1.1° (MIXS-T) FoV: 10° (MIXS-C) | Primary object is the study of the surface composition. X-ray emission from the surface will probe the electron precipitation toward the planet. Likely feasible only in the unlit surface | *PI*: E. Bunce, University of Leicester (UK) *Co-PI*: K. Muinonen University of Helsinki (Finland) |

The baseline for the MPO and Mio orbits will ensure that there are several close encounters near periherm (a few 100 km or less) throughout the mission. Such events are very important for the inter-calibration of these in situ detectors.

The identification of the optimal configurations for intercalibrating the remote sensing instruments requires a more detailed analysis. MPO/PHEBUS and Mio/MSASI will both remotely detect the Na in different emission lines, while MPO/SERENA-ELENA and Mio/MPPE-ENA will remotely detect the surface emission in ENA.

Mio/MSASI will detect the Na D lines at 589.0 and 589.6 nm and MPO/PHEBUS will try to detect the weak and challenging Na line at 268 nm. The intercalibration requires the same column density to be observed at the same time. This could be obtained when one of the two spacecraft lies along the field of view of the instrument of the other spacecraft (Fig. 13). Alternatively, assuming that the main exospheric signal comes from the near-surface regions and is symmetric and isotropic, intercalibration could be obtained when both instruments point at the same target from different point of view. The region of highest Na emission (dayside) is the best candidate for this observation.

Conversely, MPO/SERENA-ELENA and Mio/MPPE-ENA observe the 2D emission from the surface which cannot be considered isotropic, so in this case the intercalibration requires the same field of view. Since the pixels (i.e. angular resolution projected onto the surface) of the two sensors are quite different, the best configuration for the intercalibration would be when MPO and Mio are close to each other, thus both close to the periherm.

# 4 Highlights of Coordinated Observations

In this chapter some important two-spacecraft coordinated observations that will allow to obtain unprecedented results for the investigation of Mercury's environment are highlighted. A summary of the proposed observations is given in Table 4.

## 4.1 Investigation of the Interplanetary Medium at Mercury's Orbit

Mio spacecraft will be in the solar wind outside the influence of Mercury's magnetic environment during the periods of orbit near apoherm, when Mercury is in the half year of perihelion phase. In these periods, the BepiColombo/Mio Low Energy Particles (LEP: MPPE-MSA, -MIA and –MEA) and field detectors (i.e.: Mio/MGF, PWI) will be able to fully characterise the plasma environment. The observations of high energy particles and solar X-ray emission provided by Mio/MPPE-HEP, MPO/SIXS and MPO/BERM will add the assessment of the particle radiation environment and its variability over a wide energy range up to hundreds of MeV, providing the full characterisation of solar disturbances and an estimation of the GCR intensity and modulation features.

Coordinated observations between BepiColombo and other nearby space missions (distances comparable to the dimentions of interplanetary structures) that are able to monitor ambient solar wind plasma and energetic particles in a larger heliospheric context, will allow investigation of the radial expansion of the structures observed at Mercury orbit and particle transport processes in the interplanetary space, respectively. During the BepiColombo orbit phase we expect to have many suitable active missions, like SOHO, ACE, STEREO-A, Solar Orbiter and Parker Solar Probe, thus covering many radial distances and longitudinal separations (Parker Solar Probe reaching 0.045 AU and Solar Orbiter 0.284 AU and also high latitudes up to $+-33°$). For this investigation, it is particularly important to compare

**Table 4** Schematic summary of the proposed coordinated observations of BepiColombo MPO and Mio described in Sect. 4

| Scientific objective | Mercury Season | MPO condition | Mio condition | Mutual Geometric conditions | MPO instruments | MPO Instruments requirements | Mio instruments | Mio Instruments requirements | Other Observation | Section |
|---|---|---|---|---|---|---|---|---|---|---|
| Small scale processes at the magnetopause boundaries (reconnection, magnetic holes, etc...) | Spring, Summer, Autumn | At the dayside and flanks of the magnetopause | Just outside the magnetopause | As close as possible: adiacent magnetic field lines | MAG, SERENA-MIPA | high time resolution measurements | MGF, PWI, MPPE-LEP | high time resolution measurements | | 4.2 |
| Small scale processes at the magnetopause boundaries (KH instabilities ) | Winter, Spring Summer | At the dusk flanks of the magnetopause | Just outside the dusk side of the magnetopause | As close as possible | MAG, SERENA-MIPA and PICAM | high time resolution measurements | MGF, PWI, MPPE-LEP | high time resolution measurements | | 4.2 |
| Propagation of ULF waves from the KH instabilities to the inner magnetosphere | Winter, Spring Summer | Inside the dusk inner magnetosphere | Just outside the dusk side of the magnetopause | | MAG | high time resolution measurements | MGF, PWI | high time resolution measurements | | 4.2 |
| Response of magnetopause expansion or compression to solar wind conditions | Spring, Summer, Autumn | At the dayside magnetopause boundary | in solar wind | | MAG, SERENA-MIPA | | MGF, PWI, MPPE | | | 4.2 |

**Table 4** (*Continued*)

| Scientific objective | Mercury Season | MPO condition | Mio condition | Mutual Geometric conditions | MPO instruments | MPO Instruments requirements | Mio instruments | Mio Instruments requirements | Other Observation | Section |
|---|---|---|---|---|---|---|---|---|---|---|
| Induction effect after major solar events | Spring, Summer, Autumn | inside the magneto-sphere, close to the planet | in solar wind | | MAG, SIXS SERENA-MIPA and PICAM, BERM | | MGF, PWI, MPPE | | | 4.3 |
| Induction effect after major solar events | Autumn, Winter, Spring | inside the magneto-sphere, close to the planet | inside the magneto-sphere, close to the planet | different positions: like dayside, close magnetotail | MAG, SIXS, SERENA-MIPA and PICAM, BERM | | MGF, PWI, MPPE | | Other space missions could be useful for providing solar wind conditions | 4.3 |
| SEP propagation in the magnetosphere | Spring, Summer, Autumn | inside the magneto-sphere at different positions | in solar wind | | MAG, SIXS, BERM | | MGF, PWI, MPPE-HEP | | | 4.4 |
| SEP propagation toward the surface | Spring, Summer, Autumn | inside the magneto-sphere in the nightside | in solar wind | | MAG, SIXS, BERM, MIXS, SERENA-ELENA | | MGF, PWI, MPPE-HEP, MPPE-ENA | | | 4.4 |
| FTE tracing | Winter | dayside | dayside inside the magneto-sphere | same MF field line | MAG, SERENA-MIPA and ELENA | high time resolution | MGF, PWI (EWO-OFA/WFC (WPI/ MEFISTO SORBET (WPT/ MEFISTO/DB-SC), MPPE-LEP, ENA | high time resolution | Other space missions could be useful for providing solar wind conditions | 4.5 |

**Table 4** (*Continued*)

| Scientific objective | Mercury Season | MPO condition | Mio condition | Mutual Geometric conditions | MPO instruments | MPO Instruments requirements | Mio instruments | Mio Instruments requirements | Other Observation | Section |
|---|---|---|---|---|---|---|---|---|---|---|
| FTE vs external conditions | Summer | dayside | in the solar wind | | MAG, SERENA-MIPA and ELENA | high time resolution | MGF, PWI (EWO-OFA/WFC (WPT/ MEFISTO) SORBET (WPT/ MEFISTO/DB-SC), MPPE-LEP | | | 4.5 |
| Solar wind circulation around the planet seen via charge-exchange ENA | Winter, Spring | night and dusk side close to apoherm | night and dusk side | | SERENA-ELENA | | MPPE-ENA | | | 4.5 |
| Exosphere vs plasma precipitation | Winter | dayside cusps | dayside cusps | above same cusps | MAG, SERENA-MIPA, -ELENA and -STROFIO, SIXS and BERM, PHEBUS | PHEBUS before and after cusp passage | MGF, MPPE-LEP, -ENA | | Other space missions could be useful for providing solar wind conditions | 4.6 |
| Exosphere during FTE vs external conditions | Summer | dayside cusps | in the solar wind | | MAG, SERENA-MIPA, -ELENA and -STROFIO, SIXS and BERM, PHEBUS | PHEBUS before and after cusp passage | MGF, PWI, MPPE-LEP, MSASI | | | 4.6 |

**Table 4** (*Continued*)

| Scientific objective | Mercury Season | MPO condition | Mio condition | Mutual Geometric conditions | MPO instruments | MPO Instruments requirements | Mio instruments | Mio Instruments requirements | Other Observation | Section |
|---|---|---|---|---|---|---|---|---|---|---|
| He exosphere | Winter | dayside | dayside cusps | | PHEBUS | looking the cusps | MGF, MPPE-MSA | | Other space missions could be useful for providing solar wind conditions | 4.6 |
| He exosphere | Summer | dayside cusps | in the solar wind | | MAG, SERENA-PICAM, -STROFIO | | MGF, MPPE-MSA | | | 4.6 |
| Electron convection and precipitation toward the nightside | Winter | nightside | nightside | approximately same magnetotail LT | MIXS, MAG, SIXS-p | | MPPE-MEA and HEP-e and MGF | | Other space missions could be useful for providing solar wind conditions | 4.7 |
| Electron nightside convection and precipitation in relation to solar wind conditions | Spring, Summer | nightside close to periherm | in solar wind | | MIXS, MAG, SIXS-p | | MGF, PWI, MPPE | | | 4.7 |
| Dipolarization and particle acceleration in the magnetotail | Winter | inner magnetotail | far-magnetotail | approximately same magnetotail LT | MAG, SERENA-MIPA and -PICAM and -ELENA | | MPPE-MEA, -HEP-e and -ENA, MGF | | Other space missions could be useful for providing solar wind conditions | 4.7 |

**Table 4** (*Continued*)

| Scientific objective | Mercury Season | MPO condition | Mio condition | Mutual Geometric conditions | MPO instruments | MPO Instruments requirements | Mio instruments | Mio Instruments requirements | Other Observation | Section |
|---|---|---|---|---|---|---|---|---|---|---|
| Electron Stimulated Desorption signature in the nightside | Winter | nightside | nightside view of the subnadir region of MPO | | SERENA-STROFIO, -ELENA,-PICAM MIXS, MAG, SIXS-p, PHEBUS | PHEBUS observing same subnadir region before and after the passage | MSASI | looking the subnadir region of MPO | Other space missions could be useful for providing solar wind conditions | 4.8 |
| Nightside exosphere vs solar wind | Spring, Summer | nightside close to periherm | nightside | | SERENA-STROFIO, -ELENA,-PICAM MIXS, MAG, SIXS, PHEBUS | PHEBUS observing before and after the periherm passage | MSASI | looking the subnadir region of MPO | | 4.8 |
| Nightside exosphere release processes | Summer | nightside close to periherm | nightside close to periherm | | SERENA-STROFIO, -ELENA,-PICAM MIXS, MAG, SIXS, PHEBUS | PHEBUS observing before and after the periherm passage | MGF, PWI, MPPE, MSASI | MSASI before and after the periherm passage | | 4.8 |
| MIV | Late Spring | at dawn | at dawn | | SERENA-STROFIO | | MDM and PWI, MSASI | PWI in high data rate mode | | 4.9 |
| Surching for Oxydes | Spring | at dawn | at dawn | | SERENA-STROFIO, -PICAM | PICAM in high mass resolution | MDM and PWI, MPPE-MSA | MSA in high mass resolution | | 4.9 |
| 2P/Encke dust stream | late Summer | At pre-midnight | | | SERENA-STROFIO, PHEBUS | PHEBUS looing at premidnight | MDM and PWI | PWI in high data rate mode | | 4.9 |

**Table 4** (*Continued*)

| Scientific objective | Mercury Season | MPO condition | Mio condition | Mutual Geometric conditions | MPO instruments | MPO Instruments requirements | Mio instruments | Mio Instruments requirements | Other Observation | Section |
|---|---|---|---|---|---|---|---|---|---|---|
| Na exosphere dual remote observation | whole Merury's year | | | same exospheric region is observable | PHEBUS | point the target exospheric region | MSASI | point the target exospheric region | Groung-based observations | 4.10 |
| Na exospheric asymmetries | Spring or Autumn | at dawn/dusk | Mio is able to see the other hemisphere (far from the planet) | | SERENA-STROFIO | Na local density | MSASI | image the exosphere at the hemisphere not observed by MPO | Groung-based observations | 4.10 |
| 3D Na exosphere | whole Merury's year | | | MPO at the Mio limb view | SERENA-STROFIO, PHEBUS | Na local density, PHEBUS before and after | MSASI | point toward MPO | Groung-based observations | 4.10 |
| Na tail | whole Merury's year | in the night side | in the night side | | SERENA-STROFIO | observe Na exosphere in the near tail | MSASI | observe the far tail | Groung-based observations | 4.10 |
| Exospheric escape anti-sunward | whole Merury's year | in the night side | in the night side | | PHEBUS, SERENA-STROFIO | observe different elements in the near tail | MSASI | observe the Na far tail | Groung-based observations | 4.11 |
| In-situ exosphere vs low energy ions | mainly in Winter | in dayside | in dayside | | SERENA-STROFIO, -PICAM, | focus on the lowest energies of ions | MPPE-MSA, PWI | focus on the lowest energies of ions | | 4.11 |

**Table 4** (*Continued*)

| Scientific objective | Mercury Season | MPO condition | Mio condition | Mutual Geometric conditions | MPO instruments | MPO Instruments requirements | Mio instruments | Mio Instruments requirements | Other Observation | Section |
|---|---|---|---|---|---|---|---|---|---|---|
| Remote exosphere vs low energy ions | mainly in Winter | in dayside | in dayside | Mio at the limb of MPO line of sight | PHEBUS | pointing Mio | MPPE-MSA, PWI | focus on the lowest energies of ions | | 4.11 |
| Remote Na exosphere vs low energy Na$^+$ | mainly in Winter | in dayside | in dayside | MPO at the limb of Mio line of sight | SERENA-STROFIO, -PICAM, | focus on the lowest energies of ions | MSASI | focus on the lowest energies of ions | | 4.11 |
| Ionization via charge-exchange | Winter, Spring | close to the planet | night and dusk side | MPO at the limb of Mio line of sight | SERENA-PICAM, -ELENA | PICAM focusing on the lowest energies of ions ELENA observing the low latitudes ENA before | MPPE-ENA | MPO in the line of sight | | 4.11 |

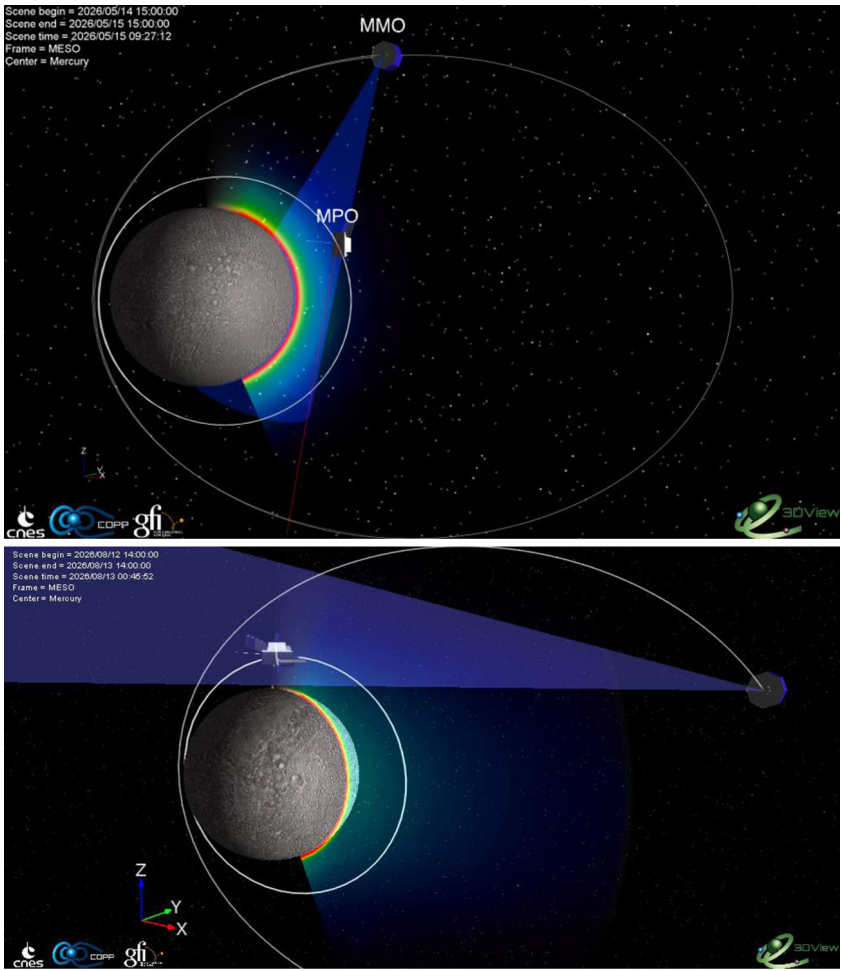

**Fig. 13**  Above: Possible configuration for intercalibrating Mio/MSASI with MPO/PHEBUS in the summer. The main emission will come from the same column density of the Na exosphere. Below: possible coordinated observation of Mio/MSASI and MPO/SERENA-STROFIO located at the closest point along the line of sight. The main emission of the Na exosphere comes from the limb, STROFIO, providing the local density, allows a 3D reconstruction of Na images recorded by MSASI

observations with modelling results of CME and solar wind propagation, SEP and GCR density gradients and propagation in the inner heliosphere.

We now highlight the most commonly used tools for this analysis. Firstly, the Inter-Planetary Scintillation 3D-reconstruction technique (IPS, Jackson et al. 2011) will provide precise tomographic 3D reconstructions of the time-varying global heliosphere and run operationally includes ICMEs and magnetic field structure. ENLIL (Odstrcil 2003) and European Heliospheric FORecasting Information Asset (EUHFORIA; Pomoell and Poedts 2018) are time-dependent 3D MHD models of the heliosphere. Both EUHFORIA and ENLIL are currently run operationally with cone CME models only (e.g., Mays et al. 2015; Scolini et al. 2018), but there are also active efforts to include flux rope CME models to obtain estimates of the detailed magnetic field structure of CMEs and their effects on plane-tary space environments (e.g., Verbeke et al. 2019). The reconstructions of the time-varying global heliosphere with IPS data will be used to iteratively update and fit ENLIL modelling (Jackson et al. 2015), to ultimately provide a rapid forecast of ICME and shocks, as well as of CIRs at Mercury. The modelling system in addition can trace the trajectories of interplanetary magnetic field lines, thereby enabling predictions of the magnetic connections from locations on the Sun to Mercury, to Earth, to BepiColombo and to simultaneously flying spacecraft. The comparison of prediction with actual measurements and timing will allow these tools to be further refined and practiced for future analyses. In fact, the efficacy of these techniques has already been tested (McKenna-Lawlor et al. 2018) to retrospectively predict the arrival of solar disturbances at Venus using timings from the Analyzer of Space Plasma and Energetic Atoms 4 (ASPERA-4) instrument suite (Barabash et al. 2007) on ESA's Venus Express spacecraft. This demonstrates that these modelling tools are able to determine the non-radial timing of solar structure arrival and the direction of high-energy particle inputs between MPO and Mio while at Mercury.

We point out that the current available models are new or have been noticeably improved since their use in support of the MESSENGER mission. As a matter of fact, prior modeling work had mainly focused on optimization of modeled results near the Earth's location at 1 AU, or at the first Lagrangian point, L1. Following studies have employed a broader range of solar wind measurements for an important validation of the models capabilities, e.g., by employing data by the dual spacecraft STEREO mission at two separated spatial locations to predict the space environment at Mercury at the time of the MESSENGER flybys (e.g., Baker et al. 2009, 2011). Moreover, the in situ MESSENGER data have been used iteratively to improve model accuracy and performance of the models for the estimation of inner heliospheric conditions. For instance, comparisons of the most recent WSA-ENLIL+Cone model results with observations by MESSENGER have indicated better predictions of the solar wind conditions at Mercury than those achieved by using WSA-ENLIL model alone (e.g., Dewey et al. 2015).

The definition of science operations of BepiColombo need to be established well in advance. Therefore, model predictions could be first used to prioritise the most interesting periods for data downloading that best illustrate the interaction of Mercury with episodes of particularly disturbed solar wind. It is now possible to perform more accurate and relatively continuous estimations of the solar wind properties near Mercury for a full exploitation of the opportunities provided by the planned investigations.

Subsequent detailed analysis of the data provided by Mio instrumentation will allow a portfolio of snapshots of the solar wind environment at different distances from the planet to be built up in response to space weather events during solar cycle 25, and to, thereby, gain important insights as to how different kinds of solar events provide and sustain different Hermean responses. Together with the models, these data will sustain an overall space weather predictions in the whole Heliosphere and a better interpretation the BepiColombo observations toward a deeper understanding of the Mercury's environment

### 4.2 Magnetospheric Boundary, Mixing Layers and Instabilities

There are several phenomena and processes acting at the Hermean magnetopause that can be investigated in new ways by two-point measurements, as well as the structure and dynamics of the magnetopause itself. With the MPO maximum altitude of around 1500-1700 km and a mean modelled magnetospheric subsolar standoff altitude of 1020 km (Korth et al. 2015), it is clear that MPO will cross the magnetopause during certain phases of its orbit (see Fig. 14). The investigation of magnetospheric boundaries requires observations of the magnetic and electric field by Mio/MGF and PWI and MPO/MAG, coupled to electron and ion observations by Mio/MPPE-LEP and by MPO/SERENA-MIPA and -PICAM.

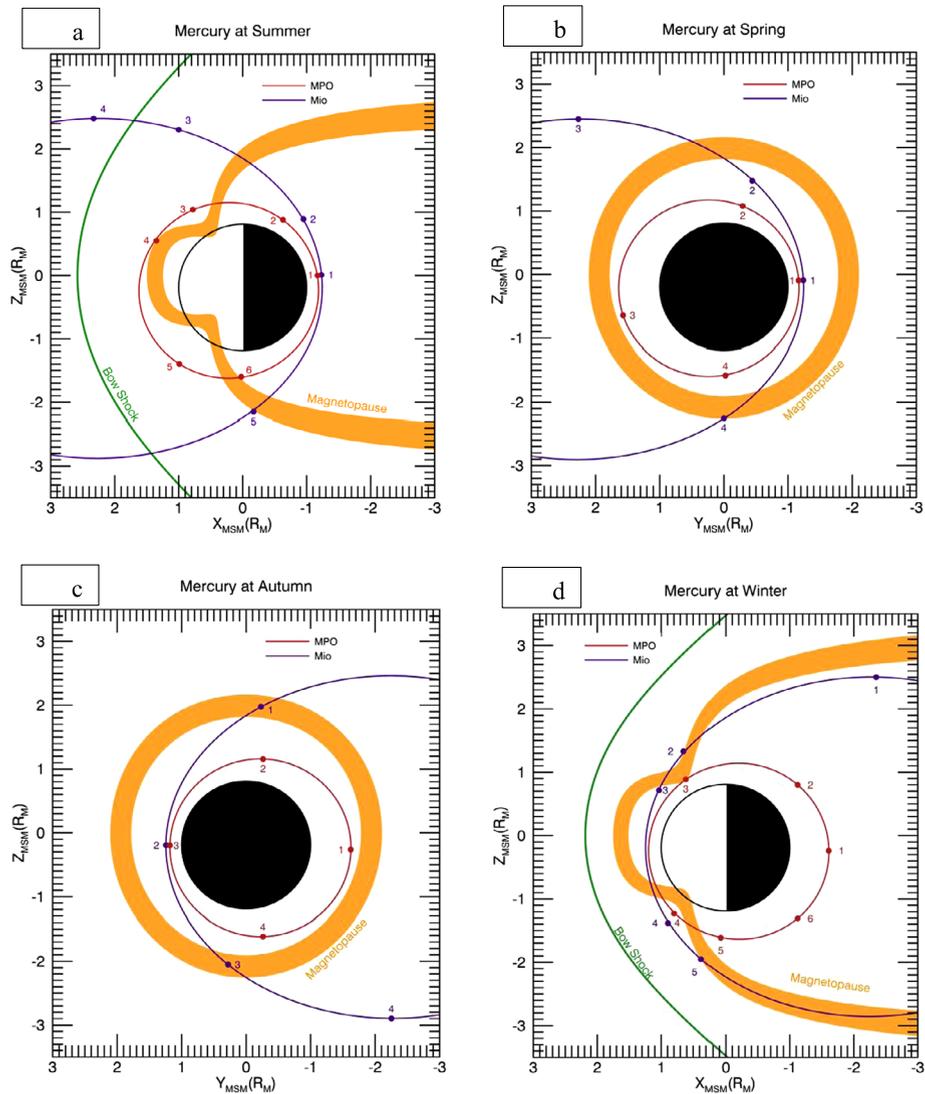

**Fig. 14** Schematic view of perihelion/Summer (**a**), Autumn (**c**), aphelion/Winter (**d**) and Spring (**b**) Bepi-Colombo orbits configurations. The planet Mercury is represented by the black circle (filled in the nightside). The red and blue lines show the MPO and Mio orbits after insertion and the dots represent spacecraft positions (*p*). The orange area represents the variability ($1\sigma$) of the magnetopause according to the 3D-model of Zhong et al. (2015a) which includes indentions for the cusp regions. The green line represents the approximate position of the bow shock (Winslow et al. 2013)

When Mio is outside the magnetosphere (see for example Fig. 14a p4), the particle and field measurements will give direct information on the solar wind properties and variations to constrain the location of the stand-off distance of the magnetopause and its dynamics. These can be measured when MPO is close to its apoherm (in the dayside, Fig. 14a between p4 and p5) by MPO/MAG and by measurement of solar wind plasma entry by MPO/SERENA-MIPA. The two-point measurements will give information on the expansion or contraction of the magnetosphere, from which we can infer the spatial scales of the magnetopause, the LLBL, and the plasma depletion layer (which is expected to have a large influence on the dayside reconnection rate; Gershman et al. 2013).

The effect of localised variations in the magnetosheath on the inner magnetosphere can be studied by two-point measurements. Such structures include large-scale pulsations associated with the cyclic reformation of the quasi-parallel bow shock (Sundberg et al. 2013), and small-scale magnetic holes (Karlsson et al. 2016). BepiColombo can investigate these structures when Mio is in the dayside magnetosheath and MPO is on the same field line at the magnetopause (Fig. 14a Mio at p3 and MPO at p4).

The processes involved in generating the LLBL at Mercury (Liljeblad et al. 2015) can be observed when Mio crosses the magnetopause flank at low latitudes (and MPO observation inside the magnetosphere can also be studied by two point-measurements, when one of the spacecraft is situated at or just outside of the magnetopause, while the other spacecraft is situated inside the magnetosphere, but close to the magnetopause, to verify the existence of magnetosheath-like plasma on closed field lines. In this case, Mio and MPO must be both close to the magnetopause. The BepiColombo orbit configurations do not include this possibility at low latitude, but coordinated observations can be performed at the polar regions (for example Fig. 14a Mio at p5 and MPO at p6 or Fig. 14d Mio at p5 and MPO at p5, or Fig. 14c Mio at p3 and MPO at p4). The outer spacecraft can then be used to evaluate plasma properties at the magnetopause to get important information about the physics at play, such as the presence of lower-hybrid waves (Treumann et al. 1991) (which can be measured by the Mio/PWI instrument), or the plasma $\beta$ as an indication of reconnection rate (Di Braccio et al. 2013). Note that at Mercury, the LLBL does not appear to be associated with KH waves, as the latter appear mainly on the dusk flank, while the LLBL typically is located on the opposite flank (Liljeblad et al. 2015) (Sect. 2.3).

As described in Sect. 2.3, the KH instability observed at the dusk flank of the Hermean magnetopause, driven by differences in plasma flow velocity and density between the two sides of the magnetopause (e.g. Sundberg et al. 2012a, 2012b, Liljeblad et al. 2014), should be investigated, coupling large scales (from two-point observations) with short scales (single spacecraft observations). This will enable a better reconstruction of the KH wave morphology (Sundberg et al. 2011) and the analysis of the influence of secondary, shorter scale instabilities driven by the primary KH instability. This can be induced magnetic reconnection or Rayleigh-Taylor instabilities, that are known to increase the mixing efficiency of the KH instability at smaller scales (Faganello et al. 2008; Henri et al. 2012). Coupled to specific coordinated observation, a fully kinetic (particle-in-cell) or multi-moment modelling approach (e.g. Deca et al. 2015; Fatemi et al. 2018; Peng et al. 2015; Chen et al. 2019; Dong et al. 2019) is required for explaining the dawn-dusk asymmetry in KH observations and for a correct evaluation of growth rates evaluation, compared to the observed KH wave activity (e.g. Nagano 1979; Glassmeier and Espley 2006; Sundberg et al. 2010; Henri et al. 2013).

MPO and Mio will be on different sides of the dusk magnetopause during Mercury seasons from Winter to Summer (Spring); the closest distances at these configurations will occur when MPO is close to the apoherm and Mio is crossing the mid-latitude magnetopause (Fig. 14b Mio at p4 and MPO at p4). Magnetic field models could help to identify measurements that are on neighbouring magnetic field lines.

The KH waves may also trigger compressional waves that may travel evanescently into the magnetosphere and mode-convert to field-line resonances (e.g. James et al. 2019). Between Winter and Spring seasons, Mio will cross the dusk magnetopause and can verify the presence of KH waves with Mio/MGF and PWI, while MPO at the apoherm is in the inner magnetotail probing the resulting ULF waves in the inner magnetosphere with MPO/MAG (Fig. 14b Mio at p4 and MPO at p3). Two-point measurements of the ULF waves in the inner magnetosphere can also give further knowledge of their nodal structure and ultimately provide information of the reflective properties of the planet surface and exo-ionosphere. Two-point measurements can also be used in a similar way to study the excitation of Alfvén waves and ion cyclotron waves by pickup ions (e.g. Boardsen and Slavin 2007) and to investigate the presence of ions energised non-adiabatically in the KH waves (Aizawa et al. 2018).

### 4.3 The Magnetosphere and Its Dynamic Response to Solar Activity

As described in Sect. 4.1, the identification of the various drivers responsible for producing space weather events in Mercury's magnetosphere will be performed by Mio in the solar wind with MGF, PWI and MPPE, complemented by MPO/SIXS observations of solar X-ray flux and BERM for radiation environment monitoring. When Mio is inside the magnetosphere and unable to monitor the solar wind conditions, we will take advantage of the MPO/SIXS-X and SIXS-P for remote observations of solar flares and SEPs and of the MPO/BERM radiation monitor as a proxy for solar wind conditions. The Sun's activity could be measured by other space missions and solar monitors operating during the Bepi-Colombo mission lifetime.

These diverse observations will then allow for identification of different types of events such as ICMEs (including their shocks and sheaths), HSS, and stream interaction regions, as CIRs. For instance, the electromotive force is found to be a significant indicator of solar-wind plasma turbulence near Mercury's orbit (Narita and Vörös 2018). Solar outbursts and ICMEs are expected to be associated with a significant peak in the electromotive force (Bourdin et al. 2018). In fact, the computation of turbulent transport coefficients and the electromotive force will be derived from the magnetic field and plasma bulk velocity observations from BepiColombo/Mio (Hofer and Bourdin 2019). Simultaneous observations of the downstream magnetic field by MPO/MAG will allow discrimination of the external and internal contribution to the magnetic field and the induction effects at the planetary core (as explained at Sect. 2.2). Magnetic field observations to study induction should be made as close as possible to the dayside mid-latitude of the planet surface, where the sub-surface currents are expected to circulate. In Fig. 14d, we can see that the closest MPO position at the mid latitude dayside is during Winter (p3 and p4) when Mio is never in the unperturbed solar wind. In this case, the solar wind measurement by different space missions could be a great support. On the other hand, in Summer (Fig. 14a) the occurrence of a solar event when MPO is in p3 and Mio is in p4 will offer the unique opportunity of unambiguous discrimination of internal and external contribution for the magnetosphere configuration with the coupled measurement of MPO/MAG and Mio/MGF and PWI.

Simultaneous observations from the MPO instrument suites (MPO/MAG, MPO/SERENA -MIPA and -PICAM) will enable snapshots of the environment at different near-planet locations; i.e.: dayside compressed magnetosphere in the closed-field-line region or below the cusps (Fig. 14d MPO between p3 and p4), the magnetospheric dawn and dusk flanks (Fig. 14b and 7c MPO p1 and p3 and Mio p2 in Autumn and p1 in Spring), the night-side mid latitude magnetosphere at closed field lines (Fig. 14a MPO and Mio at p1 and p2), the near tail (in Winter Fig. 14d MPO at p6, p1 and p2), far tail (in Winter Mio will be able to *sound* for the first time the far tail response to solar events Fig. 14d between p5 and p1). Finally, it will be possible to deduce how Mercury interacts over time with the varying solar wind at a range of altitudes. The evaluation of the fraction of solar wind impacting the surface could be used as an indirect measurement of the magnetic field strength at the surface (see also Sect. 4.5).

## 4.4 The Effect of SEPs on the Magnetosphere

When SEPs impact Mercury, a significantly enhanced flux of energetic particles will reach the surface, possibly contributing to planetary gamma ray-emission, to X-ray fluorescence and even to exospheric changes (e.g. Potter et al. 1999; Leblanc and Johnson 2003), depending on the SEP features and IMF orientation (Sect. 2.5).

Direct observations of SEPs will be obtained by MPO/SIXS-P to determine the characteristics of individual SEP events (e.g. spectra, fluence, maximum flux, anisotropies) over a wide range of energies ($\sim$1–30 MeV for protons) and to establish the range of SEP variability from event to event at Mercury's location. Simultaneous observations of Mio/MPPE-HEP and MPO/BERM will complement MPO/SIXS to extend the energy range in the low and high part of the spectrum, respectively, when Mio and MPO are close to each other or both in the magnetosheath (Fig. 14a MPO at p5 and Mio at p3). Comparison of these observations with SEP observations from spacecraft elsewhere in the inner heliosphere (Solar Orbiter, Parker Solar Probe) will allow for unprecedented multi-spacecraft studies of SEP transport in the innermost region of the heliosphere.

On the other hand, when the two spacecraft are more distant from each other, i.e.: Mio in the solar wind and MPO closer to the planet in the magnetosphere (Fig. 14a Mio at p4 and MPO at p2), comparison between Mio/HEP and MPO/SIXS at energies around 1 MeV (100–700 keV) for protons (electrons) can provide info about the particle access in the magnetosphere at such energies, for which the particle trajectories are more affected by the magnetic field.

In addition, the combined use of Mio/HEP and MPO/SIXS observations in different spacecraft configurations coupled to modelling of the SEP interaction with the Hermean environment will allow a more accurate investigation of the SEP propagation in the magnetosphere to estimate the motion of charged particles inside the magnetosphere (e.g., the presence of sustained trapped particles). It will be possible to evaluate the SEP fluxes reaching the surface and their role in inducing fluorescence, by taking advantage of simultaneous measurements of MPO/MIXS, especially when MPO is located in the night-side region, thus avoiding the otherwise dominant contribution from solar X-ray fluorescence and scattering. Finally, although the SEPs could produce a higher background noise level within the sensors, in principle, coordinated and simultaneous measurements by MPO/SERENA-ELENA and the Mio/MPPE-ENA could be able to identify possible back–scattering from the surface while SERENA-STROFIO and the exospheric remote sensing instruments MPO/PHEBUS and Mio/MSASI will allow investigating a potential SEP contribution to the exosphere.

## 4.5 Magnetospheric Ion Circulation and Acceleration Processes: Dayside Flux Transfer Event

Flux ropes are reconnection-related magnetic structures which are observed to propagate away from reconnection sites on the dayside magnetopause and in the magnetotail. They accelerate ions and electrons and can lead to particle precipitation onto the planetary surface (e.g., Raines et al. 2014) (see Sect. 2.3). Detailed studies using MESSENGER data investigated the duration and diameter of dayside flux ropes frequently observed in the cusps and near local noon (known as FTEs) to be ~2 s and 0.078–0.52 $R_M$, respectively (e.g., Slavin et al. 2008, 2009, 2010; Imber et al. 2014; Leyser et al. 2017).

The BepiColombo Mio/MGF and MPO/MAG high time resolutions (up to 128 Hz) will be coupled to high time resolution (up to 4 s, i.e. the Mio spin period, for a full energy and angular coverage) of the Mio/MPPE ion and electron sensors, MIA, MEA and MSA, and to a good time resolution (up to 20 s for a full energy and about half sphere angular coverage) of the MPO/SERENA ion sensors (Table 3). In comparison with the MESSENGER measurements these performances will allow a much deeper analysis of the FTE structure and evolution as well as of the acceleration processes inside the magnetosphere.

First of all, BepiColombo will be able to trace bursts of FTEs (known as FTE showers) and may indirectly observe the related surface precipitation by the detection of the directional neutrals coming from the surface. In Winter, where both Mio and MPO will be located in the dayside magnetosphere, FTEs showers will be observed by both spacecraft located near the same field line (Fig. 14d MPO at p3 and Mio at p2). Ion precipitation related to FTEs is a major topic of scientific interest evidencing Mercury's interaction with the solar wind. FTEs can be identified by MPO/MAG, while MPO/SERENA-MIPA will detect ion fluxes toward and away from the planet, while SERENA-ELENA will detect back-scattered particles from the surface, indicative of ion flux impact onto the surface. Mio will provide a third sampling point providing full ion angular distributions to trace the particle trajectories with MPPE-MIA and a second measurement of back-scattered particles from the surface with MPPE-ENA. The coupled measurements of the magnetic and electric fields by two magnetometers and PWI will enable the determination of particle trajectories. Since the duration of each FTE is comparable to the time resolution of MPPE or relatively short compared to that of detectors onboard MPO, it will be difficult to know the influence of each FTE on particle precipitation. However, many cases of multiple flux ropes can be expected to be observed, thus the ion precipitation due to FTEs can be investigated. Especially, we expect to examine how much energy can be transported by FTEs and related cusp filaments (e.g., Poh et al. 2016; Raines et al. 2014) to the exosphere and/or surface.

A second coordinated observation will allow assessing the influence of the solar wind properties on FTE characteristics (e.g., how broadly are FTEs observed, which parameter is important to determine the size of FTEs). In Summer, while Mio observes the solar wind (as described in Sect. 4.1), MPO located near the dayside cusp region (Fig. 14a MPO between p3 and p4 and Mio at p3) will make the measurements outlined in the previous observation. The investigation of FTE influence on the exosphere will be discussed in Sect. 4.6.

The third coordinated observations will allow BepiColombo to follow the solar wind trajectories inside the magnetosphere. Those solar wind particles, entered inside the magnetosphere, not impacting the surface and not fully mirrored toward the Sun, could be transported westward toward the nightside passing close to the planet at dawn magnetospheric flank (Mura et al. 2005, 2006). Charge-exchange ENA generated by the interaction between the exosphere and the protons circulating close to the planet surface (see Sect. 2.8) can be detected from the nightside in Winter or duskside in Spring looking toward the planet by both MPO/SERENA-ELENA and Mio/MPPE-ENA (Fig. 14b MPO at p3 and Mio between p3 and p4 or Fig. 14d MPO at p1 and Mio between p5 and p1). The lateral sectors of the SERENA-ELENA FoV will point the East limb when MPO will be at the apoherm, hence it will be able to register the charge-exchange ENA circulating westward at low latitudes. Instead, MPPE-ENA will provide a global view of charge-exchange ENA. These observations will provide for the first time the remote sensing the close-to-planet plasma circulation in ENA imaging at Mercury. Reconstruction of the 3D ion distribution or the plasma trajectories will be obtained by using ENA deconvolution techniques (e.g.: McComas et al. 2009) that greatly improve when double observations are available, that is when both spacecraft will be in the night or dusk apoherm.

## 4.6 Dayside Exosphere Response to Solar Activity

While MESSENGER observations of different exospheric species were not able to register a strict connection of the Sun's activity to the exosphere morphology or intensity, as described in Sect. 2.7, the frequent two-peak pattern observed by ground-based telescopes in the Na exosphere at dayside mid latitudes is considered to be strictly related to the solar wind precipitation at the magnetic cusp (e.g.: Mangano et al. 2013, 2015; Massetti et al. 2017, Potter et al. 2006), hence related to dayside FTEs, as considered in Sect. 4.5. Moreover, rapid changes into a single-peak pattern in the equatorial sub-solar region (Orsini et al. 2018) could be the signature of morphology changes induced by impulsive solar events like ICME. Nevertheless, the exact mechanism responsible for such surface release is still under discussion, since there are many possible processes (like ion-sputtering, chemical sputtering and PSD) that act simultaneously and influence each other (e.g.: Mura et al. 2009) (see Sect. 2.7).

The only way to unambiguously solve the question requires simultaneous measurement of solar wind and IMF, of precipitating ions, of backscattered particles (proving the impact onto the surface), in-situ measurement of exospheric component variations and/or remote sensing of exospheric distributions and vertical profiles. After a major precipitation event, the action of ion sputtering process alone would cause a density increase for almost all the exospheric components in an energetic distribution, while the action of PSD enhanced efficiency, following ion impact, would cause a density increase for only the volatile components. BepiColombo will offer different possible coordinated measurements configurations able to accomplish these requirements.

At the aphelion phase (Winter), when both Mio and MPO orbits have the periherm in the dayside, they will be above the cusps at different positions (Fig. 14d MPO at p3 and Mio between p2), the Solar wind activity could be monitored by MPO/SIXS and MPO/BERM, while the fluxes of the precipitating particles will be measured by the ion spectrometers MPO/SERENA-MIPA and Mio/MPPE-MIA. The back-scattered particles from the surface will be monitored by SERENA-ELENA and MPPE-ENA, providing the map of the precipitation. SERENA-STROFIO in situ measurements will register any fast variation of the different exospheric components, looking for time-shifted relation with the precipitating fluxes. MPO/PHEBUS can observe the exospheric vertical profiles at mid latitudes before and after the cusp passage looking for short-time variability of different species and variation of the scale height (proxy of the energy distribution).

At perihelion phase (Summer), when both Mio and MPO orbits have their periherm in the nightside (Fig. 14a MPO at p3 and Mio at p4), Mio will be frequently in the solar wind thus allowing a detailed characterisation of the solar activity, as described in Sect. 4.1, with the support of MPO/SIXS and BERM. MPO above the cusps will be able to register the solar wind precipitation with MPO/SERENA-MIPA and impact onto the surface with SERENA-ELENA. At the same time, if the geometry allows it, Mio/MPPE-ENA can add a more global detection of backscattered particles from the planet surface. As in the previous case, SERENA-STROFIO in-situ measurements will register any fast variation of the different exospheric components and MPO/PHEBUS can observe the exospheric altitude profile before and after the cusp passage. In this case, Mio/MSASI will provide an additional global image of the Na exosphere.

Both these coordinated observations could be coupled to the coordinated measurements described in the previous Sect. 4.5 adding the investigation of the exosphere reaction to FTE occurrence.

Such observations should be performed under different solar conditions, i.e.: nominal solar wind conditions CMEs, HSS, and CIRs, to account for the different responses of the exosphere. This implies that particular care in data download selection is required. As mentioned in Sect. 4.1, other spacecraft in the inner Solar system as well as specific space weather forecasting or disturbance propagation modelling tools could be used to support such investigation.

As introduced in Sect. 2.6, the He exosphere offers several points of interest related to gas interaction with the surface and identification of endogenic and radiative decay of the surface material. The detailed ad simultaneous measurement of neutral He by PHEBUS and SERENA-STROFIO on MPO, and alfa particles of the solar wind by MPPE-MSA on Mio and SERENA-PICAM on MPO will allow investigating if neutral and ionised components are related to each other or if there is a He component unrelated to solar wind, hence due to an endogenic source, as in the Moon case. This simultaneous measurement can be achieved when PHEBUS can observe the dayside exosphere and Mio/MPPE-MSA is inside the dayside magnetospheric cusps (Fig. 14d MPO at p2 and Mio at p3) or when MPO/SERENA-PICAM is inside the dayside magnetospheric cusps together with SERENA-STROFIO observing the ionised and neutral He components, respectively, while Mio/MPPE-MSA is in the solar wind observing the alfa particles (Fig. 14a MPO between p3 and p4 and Mio at p4).

## 4.7 Magnetospheric Ion Circulation and Acceleration Processes: Magnetotail Dipolarisation and Convection

X-ray emission observed from Mercury's nightside surface by MESSENGER's XRS was not optimal for the characterisation and mapping of the precipitating population. In particular, MESSENGER's elliptical orbit did not allow all regions of the surface to be equally accurately characterised (see Sect. 2.4).

MPO/MIXS will be able to detect electron-induced X-ray emission from the surface. The high energy and spatial resolution of MIXS, along with the north-south symmetry of the MPO orbit, will allow improved characterisation of the regions of X-ray emission, although operational and background signal constraints make these measurements simpler for unlit regions of the surface.

The measurement could be further improved through coordinated measurements with other instruments on MPO and Mio. While MIXS observes this X-ray emission, other instruments on both spacecraft will collect complementary in situ particle and magnetic field data (MPO/SIXS, MPO/MAG, Mio/MPPE-MEA and -HEP-e). While precipitation at all nightside magnetic local times has been observed by MESSENGER, the emission is most frequent in the MLT sector 0–6 hours, thus we expect to see enhanced emission during the mission phases when the MPO and Mio orbital planes span these local times. Two possible configurations are optimal; in Spring toward Summer, MPO periapsis on the nightside and Mio apoapsis on the dayside in the solar wind will provide the highest spatial resolution at the locations of interest while monitoring the solar wind conditions (Fig. 14b MPO at p1 and Mio at p3 or Fig. 14a MPO at p1 and Mio at p4), while in Winter, apoapsis on the nightside and periapsis on the dayside will allow Mio to operate as a down-tail monitor, observing reconnection-related magnetic field and plasma events in situ. Mio's apoapsis will remain within the magnetopause for MLTs between ~22 h and 2 h; optimal conditions will therefore occur at 0–2 h MLT (Fig. 14d MPO at p1 and Mio between p5 and p1).

While the nominal operating configuration of MIXS is sufficient to investigate electron-induced X-ray fluorescence in detail, it may be possible to operate the instrument to lower the energy detection threshold, allowing detection of additional fluorescence lines, and to increase the MIXS-C pixel rate, allowing improved time resolution (see Bunce et al. 2020, this issue). Either or both of these configuration changes would significantly increase the MIXS data rate, so they must be restricted to the scenarios in which they can be of the greatest benefit.

As stated in Sect. 2.4, Mercury's magnetotail observations have demonstrated that reconnection signatures may be routinely observed by a spacecraft passing through the magnetotail at a downtail distance of 1-4 $R_M$. (Imber and Slavin 2017; Sundberg et al. 2012a, 2012b; Smith 2017; Dewey et al. 2017). Mio's orbit will be most favourable for observations of tailward flux ropes in Winter, when Mio's apoapsis is on the nightside, encountering the equatorial magnetotail at a distance of 4 to 6 $R_M$. The relevant instruments for this observation will be Mio/MPPE, -MGF, and PWI. The MPO observations in the near tail of magnetic field and ion fluxes by MPO/MAG and MPO/SERENA ion spectrometers will confirm the transit of planetward dipolarisation fronts and accelerated particles (Fig. 14d MPO at p1 and Mio between p5 and p1). Observations made by MESSENGER were unable to directly investigate causal links between these processes and the electron-induced X-ray emission, which was also predominantly observed in the dawn hemisphere. Joint observations by MPO and Mio will enable simultaneous observations in both the mid-tail and near-planet region, and hence greatly improve the characterisation of the links between these regions.

Additionally, the in-situ observations of planetward moving plasma and nightside backscattered particles observed by MPO/SERENA-ELENA and Mio/MPPE/ENA will enable an assessment of the extent and significance of precipitating ion populations. This will enable a comprehensive analysis of the source and nature of the particles that precipitate onto the surface and contribute to analysis of exospheric source processes (see Sect. 4.8).

### 4.8 Exosphere Response to Nightside Particle Precipitation

Precipitation of electrons onto the surface is a potential exospheric source through ESD (Domingue et al. 2014) (see Sects. 2.4 and 2.5). Through coordinated observations of precipitation to the surface by MPO/MIXS and observation of exospheric species either at the same time or under similar viewing conditions by MPO/PHEBUS, Mio/MSASI and SERENA-STROFIO, BepiColombo will help to establish whether this phenomenon contributes significantly to the Mercury's exosphere.

ESD can result in the release of alkalis (Na and K) compounds, alkaline earths (e.g. Ca and Mg, Bennett et al. 2016) and ions (e.g. McLain et al. 2011). Observations from MESSENGER XRS (Lindsay et al. 2016) imply that electrons of sufficient energy to induce Mg-K$\alpha$ and Na-K$\alpha$ fluorescence regularly precipitate to the nightside surface; electrons with sufficient energy to fluoresce Ca-K$\alpha$ also precipitate, although less frequently. MIXS could be able to detect the Na-K$\alpha$ fluorescence (at 1.04 keV), fulfilling two purposes: the detection and quantification of relevant species on the surface as potential sources (in this case simultaneity is not required), and the location of areas of enhanced particle precipitation on to the surface through detection of electron-induced X-ray fluorescence.

The MPO/PHEBUS and Mio/MSASI remote sensing of the exosphere will register possible volatile species release subsequent to electron precipitation events. If the electron precipitation is registered during MPO periherm passage, MPO/SERENA-STROFIO will add in situ observation of variability of volatile species density with some delay due to the exospheric transport at the MPO altitudes.

Nightside observation is preferred to eliminate PSD as a competing exospheric source and for reducing background signals due to photon-induced fluorescence. Favourable geometries for MPO observations are similar to those described in Sect. 4.7, i.e. with the orbital plane between 0 and 6 hours local time. While Mio/MSASI should be able to observe the exosphere at the same time with a field of view including the MIXS footprint, ideally near to the limb. This coordinated observation could be performed in Spring toward Summer, MPO close to periapsis on the nightside and Mio in the nightside (Fig. 14b MPO at p1 and Mio at p2 or Fig. 14a MPO at p1 and Mio at p5 or p2), or in Winter, MPO and Mio in the nightside (Fig. 14d MPO between p6 and p2 and Mio at p1).

MPO spacecraft geometry means that simultaneous observations of this nature by MIXS and PHEBUS are not possible; instead, observations must be designed with the shortest possible gap between a MIXS surface observation and a PHEBUS observation of the exosphere at the same local time. Observations from MIXS during the mission will allow us to develop a more complete understanding of the stability of electron precipitation with time.

In the Summer nightside-periherm phase (Fig. 14a MPO at p1 and Mio at p1), together with electron precipitation mapped at the surface by MPO/MIXS, MPO/SERENA-PICAM and Mio/MPPE-MSA will be able to infer (with MPO/MAG and Mio/MGF support) the ion precipitation toward the surface, while and MPO/SERENA-ELENA and Mio/MPPE-ENA will map possible back-scattering from the surface. As in the coordinated observations discussed above MPO/SERENA-STROFIO is a good candidate for detecting a signature of exosphere response considering the necessary delay time due to exospheric transport to the spacecraft altitude (about 10s minutes, Mangano et al. 2007), Mio/MSASI can add Na imaging while observing the precipitation region at the limb of its field of view (Fig. 14a Mio at p5 or p2) and MPO/PHEBUS observations can be added with a delay time due to pointing constraints, as described above.

### 4.9 Exosphere Response to Micrometeoroid Impacts

Mio/MDM is designed to detect the impact of momentum and velocity as well as the concentration of micrometer-sized grains. The sensor will detect the incoming dust particles as well as the ejecta cloud released from the surface. The discrimination between these two populations will be done by an a-posteriori analysis of the data (Nogami et al. 2010; Kobayashi et al. 2020, this issue). Furthermore, Mio/PWI uses its four antennas to measure the surrounding electric field. High resolution data down-link will allow use of these antennas to detect the impact of micrometre-sized grains on either the spacecraft body or the electric field antenna itself. Being the dust grain charged, the coordinate observations of these dust detectors together with the magnetometer on Mio are important to determine the dust trajectories down to the surface. The plasma particle observations will help to study the complex dynamics of the small dust grains; in fact, their charging depends on the surrounding plasma (density and energy) and the motion can be affected by the electric field induced by the solar wind plasma flow, especially in an extreme solar wind condition like CME (Czechowski and Kleimann 2017). The exospheric response to micrometeoroid impacts can be investigated by searching for refractories or atom groups (generally signature of MIV or ion sputtering surface release processes) by MPO/SERENA-STROFIO locally and by MPO/PHEBUS remotely. Simultaneously, Mio/MSASI could add the Na global imaging for investigating the effect of MIV on Na distribution. This set of observations will allow the characterisation of the released material, composition and vertical profile in relation to micrometeoroid input. Thus, an estimate of the properties of the vaporised cloud, like quenching temperature, dissociation and photolysis lifetime generating Ca, Mg, Na and K atoms from the their oxyde in the Hermean exosphere will be obtained (Sect. 2.8).

Identifying molecules in Mercury's exosphere will also help anwer the question about the oxygen fate at Mercury formulated in Sect. 2.8. The combined observations of SERENA-STROFIO mass spectrometer that will be able to discriminate the neutral particle masses and SERENA-PICAM (Orsini et al. 2020, this issue) and MPPE-MSA (Saito et al. 2020, this issue) discriminating the charged particle masses will allow to identify many oxydes, although in some cases partices with the same masses (as $O^{17}$ and $O^{18}$ with OH and $H_2O$, respectively) could be confued (for exemple see Fig. 14b MPO at p1 and Mio at p1).

If micrometeoroid impacts occur mainly in the ram direction, that is, in the dawn hemisphere, the best occasion to observe the MIV cloud would be in Spring (Fig. 14b) when both spacecraft have periherm at dawn (MPO and Mio close to p1). Whatever the case, observations close to the shadowed surface (the perihelion half year, late Spring) by MPO/SERENA-STROFIO will be particularly important to investigate the MIV process detecting refractories and atom groups, since the surface release due to other processes like PSD and thermal desorption are not active (see Sect. 2.8).

As described in Sect. 2.6 and 2.8, at TAA about 30°, while crossing the comet 2P/Encke dust stream, the Ca column density increases (Killen and Hahn 2015). According to Christou et al. (2015), the dust stream impact onto the surface is in the dusk side before midnight, the exospheres of refractories and their oxydes, like Ca and CaO or Mg and MgO, are expected to increase in the same region, but observable by UVVS spectrometer only when rising in the dawn side (Plainaki et al. 2017). The periherm of Mio and MPO will be specifically at pre-midnight in that period (See Fig. 12, TAA 45°), so that STROFIO will be able to register a possible signal to give some context to the MDM dust detection. Furthermore, MPO/PHEBUS will be able to remotely sense the Mg exosphere reaching the illuminated side.

## 4.10 The Special Case of the na Exosphere

The study of the exosphere of Mercury is one of the main goals of the BepiColombo mission and, among the species that constitute the exosphere, sodium is certainly one of the most abundant and the most observed by space-based and ground-based observations. Despite the numerous observations of the Na exosphere from ground-based telescopes and space-based spectroscopes, its source, distribution and variability along Mercury's orbit, with local time and with latitude, as well as its relationship with surface composition, cannot be explained in a comprehensive scenario (see Sect. 2.8). In fact, several processes, such as PSD, ion sputtering, MIV, ESD or direct thermal release, have been suggested as surface release mechanisms. Even including the action of radiation pressure and the photoionization, all the observed features cannot be explained unequivocally. The relative importance of these processes for Na should be addressed with the help of theoretical models and constrained with in-situ data, since direct access to the location of the source (Mercury's surface) is not possible. Na exosphere investigation is a fundamental tool of analysis to study all other species, since the study of sodium sheds light on all possible sources and sinks of the exosphere. In this respect, BepiColombo offers a comprehensive payload for the investigation of Na exosphere, its possible drivers, the circulation mechanisms, and its fate. The instruments devoted to exospheric investigations on MPO (SERENA-STROFIO and PHEBUS) are coupled with Mio/MSASI that is specifically designed to image sodium. Coordinated observations of ion and electron precipitation by MPO/SERENA and MPO/MIXS and Mio/MPPE detectors (see also Sects. 4.6 and 4.8) and the dust monitoring by Mio/MDM (See Sect. 4.9) will be compared to the Na exosphere obtined by simulating the surface release processes and the exospheric mechanisms with different input parameters, like velocity distributions, surface mineralogy, yield, radiation pressure, etc.. (i.e. Mura et al. 2007; Gamborino and Wurz 2018). Eventually, it will be possible to unambiguously determine the contribution of each driver.

The first, obvious, coordinated observation is the remote-sensing of the same portion of exosphere with different sensors on board MPO and Mio, such as PHEBUS and MSASI, respectively. Even if, the PHEBUS observation of the Na line at 286 nm is expected to have a high background noise (Quèmerais et al. 2020, this issue), such observation could be performed from different vantage points, to obtain an averagespatial distribution. In most cases, MPO and Mio observe Mercury from different locations, allowing the strategy described above; in the rare event of MPO and Mio being close to each other, it would be appropriate to observe the same line of sight to cross-calibrate the two instruments (Sect. 3.3). During several phases of the mission, it would be possible for these instruments to simultaneously observe the dayside, possibly above the cusps, which are important regions for the release of many exospheric species and where the Na is expected to be intense enough to be detected by PHEBUS.

Alternatively, a more intriguing scenario involves the use of an in-situ experiment on board MPO (SERENA-STROFIO) and a remote-sensing instrument on board Mio (MSASI) to detect Na exosphere. In this case, MSASI should observe the column density of sodium along the line of sight that includes MPO at the limb (closest point to the surface), while SERENA-STROFIO should measure the sodium density in the supposed highest density contribution for the column density (Fig. 13). Such a data-set could then be used to retrieve the 3D density of sodium, a task that is quite critical.

Observing the same location with two instruments is not the only way to exploit a two-spacecraft mission. The intense time variability of the Na exosphere of Mercury makes it possible to observe sporadic events of intensification of sodium and to follow them around the planet looking for asymmetries and variations. The dawn/dusk asymmetries can be investigated by observing with SERENA-STROFIO the Na exosphere at MPO orbit in Spring or Autumn (periherm is the best case) while Mio/ MSASI will image the other hemisphere (Fig. 14b MPO at p1 and Mio at p3 or Fig. 14c MPO at p3 and Mio at p4). The Na intensification could be traced from the release region, which could be observed by one sensor (MPO/SERENA-STROFIO), down to the exospheric tail easily observed by Mio/MSASI, which is generated by the action of the radiation pressure where a large part of sodium is lost.

In particular, the simultaneous observation of the Na variability at the terminator, comparing the Na-exosphere surface densities in the Sun-light and in the shadow would allow the investigation of the effect of PSD and of the surface temperature on the Na release.

Finally, sodium can be observed from several Earth-based telescopes in the visible (THEMIS, McDonald Observatory, AEOS, Vacuum Tower Telescope). Hence, it is desirable to plan ground-based observation campaigns in conjunction with BepiColombo observations. The same strategies described above for two spacecraft apply in the case this third vantage point joins the coordinated observations. In other words, the Na exospheric global distribution, its variability and dynamics will be analysed with great detail highly improving the science return.

## 4.11 Exosphere Sinks and Planetary Loss

In a surface-bounded exosphere the surface is both the source and the main sink for the exospheric particles. If a neutral particle does not precipitate back onto the surface, this is because it is ionized (after collision with a photon or with some other particle), dissociated, or lost to space by Jeans escape or the radiation pressure effect (see Sect. 2.4). The newly ionized particle is quickly accelerated and partially mixed with the down-streaming solar wind (see Sect. 2.3). Estimation of the planetary loss is crucial for investigation of Mercury's evolution as well as for constraining models of evolution of exoplanets close to their parent star (Mura et al. 2011). Under different conditions of illumination, radiation pressure, and interplanetary medium encountered along the highly eccentric orbit of Mercury, and depending on the considered component, different loss processes could play the dominant role. Investigation of the exospheric sinks could be performed by BepiColombo by different coordinated observations.

The direct loss of neutral components can also be measured by observing the tail generated by acceleration of specific species due to the radiation effect that is proportional to the radial heliocentric velocity of the exospheric atoms; such velocity depends on Mercury's trajectory (Schmidt et al. 2012). The most evident effect of the radiation pressure is the observed Na tail that is highly variable along Mercury's orbit (see Sect. 2.6). BepiColombo will be able to observe the Na tail with great details thanks to Mio/MSASI; in fact, during expected denser tail periods, i.e. during mid seasons, the Na tail can be imaged in nominal operations up to 25 $R_M$, and even more during specific observations campaigns. Furthermore, thanks to MPO/PHEBUS it will be possible to map the tail of other species affected by radiation pressure, like K, to be compared to the Na tail. Ground-based observations specifically targeted to the tail could add a further support to the investigation.

The simultaneous observations of the same species in the form of low-energy ions and of neutral component at the dayside (for example in Winter, Fig. 14d MPO and Mio between p3 and p4) would allow estimation of the link between the two populations and hence of the photo-ionization rate. MPO/SERENA-STROFIO and SERENA-PICAM will be able to observe simultaneously and at the same point diverse exospheric components and low energy ions down to few 100 eV (depending on spacecraft potential). When Mio will be close to MPO, Mio/MPPE-MSA will provide the low energy population to be compared to the ones observed by PICAM. Thanks to Mio/PWI it will be possible to evaluate the Mio spacecraft potential thus reaching a better estimation of the lower-energy ion density.

In the situation when MPO is far from the planet and the PHEBUS line of sight includes Mio at the limb (for example, Fig. 14d MPO at p3 or p4 and Mio at subsolar point), this will allow the exospheric variations of diverse species to be compared with the Mio/MPPE-MSA observations of the low-energy ion components. Similarly, when MPO is in the line of sight of MSASI at limb, SERENA-STROFIO will provide the local Na density while SERENA-PICAM will measure the low-energy $Na^+$.

Furthermore, the charge-exchange ENA observations, as described in Sect. 2.7 and 4.5, are a signature of exosphere ionization. These ENA, generated by the exosphere interaction with the circulating solar wind close to the planet, will be detected by line-of-sight pointing at the limb, so they can be easily discriminated from back-scattered ENA coming from the planet surface. They exit the interaction region leaving behind low energy ions, so that, simultaneous remote sensing of ENA at the limb and of in-situ measurement of low energy planetary ions is a way to simultaneously observe the ion circulation, the exospheric loss and its fate. This measurement is feasible for BepiColombo by observing with Mio/MPPE-ENA when the apoherm is in the nightside or dawnside and MPO/SERENA-PICAM is in the ENA line-of-light limb (for example Fig. 14d MPO between p2 and p3 and Mio in the tail). The SERENA-ELENA sensor will be able to discriminate the charge exchange ENA at the low latitudes, observing the dawn/night limb from the apoherm during winter/spring. This measurements could preceed and complement to the previous measurements (Fig. 14d MPO at p1).

Finally, BepiColombo will have the chance to constrain the loss rate for $^{40}$Ar. In fact, MESSENGER MASCS bandpass did not include the emission line doublet of $^{40}$Ar at 104.8 and 106.7 nm, but PHEBUS bandpass does. A measurement of the column density of $^{40}$Ar (and hence of its source rate) would constrain the abundance of $^{40}$K within the crust, with important implications for Mercury formation. However, instantaneous detection by PHEBUS of neutral $^{40}$Ar will be challenging, because of its extremely low resonant scattering efficiency, and will likely require integration of spectra taken over multiple orbits (Quèmerais et al. 2020, this issue). Finally, SERENA-PICAM and MPPE-MSA could measure the flux of escaping ions ($^{40}$Ar$^+$) being the photo-ionization and electron impact ionization the major loss processes for this species.

## 4.12 Surface Observation as a Support for the Environmental Investigations

The MPO instrument suite will map the elemental and mineralogical composition of the surface, with equally good spatial resolution in northern and southern hemispheres. MIXS will be sensitive to more elements than MESSENGER's instrument and will be supplemented in both hemispheres by additional elemental detections by the gamma ray and neutron spectrometer component of MPO/MGNS. The mappable elements (at diverse spatial resolutions) include Si, Al, Fe, Mg, Ca, S, Ti, Cr, Mn, Na, K, P, Ni, U, Th, Cl, O, H and possibly C (for further discussion of elemental mapping by MIXS and MGNS see Rothery et al. 2020, this issue).

Mineralogic information will come from the thermal infrared spectrometer (MPO/MERTIS – Hiesinger et al. 2020, this issue) – a type of instrument that has never been used at Mercury before – and also from the visible and near infrared spectrometer (MPO/Simbio-Sys-VIHI – Cremonese et al. 2020, this issue). For a better investigation of the link between the surface and the exosphere, the observations of the exosphere should be as close as possible to the surface region of interest.

The availability of composition and mineralogy maps will allow the investigation of the preferential release process active at specific targets, like hollows or polar deposits (see Sect. 2.9). The surface characterisation will allow a better estimation of the parameters required for modelling the surface release processes (i.e. yields, species concentrations, etc. . . ) and the generated exosphere. Observations of the close-to-surface volatile component by PHEBUS UV spectrometer coupled with the in-situ measurement of the SERENA–STROFIO mass spectrometer and the MSASI low altitude Na distribution will constrain the exospheric modelling for the identification of the active surface release process and release rate.

Conversely, the average refractory versus volatile distribution of the exosphere, obtained by MPO/SERENA-STROFIO and PHEBUS, weighted by the surface composition mapping will provide information on the drivers of the surface release. The expected drivers for the release of the refractories are micrometeoroid and ion impacts; hence, the weighted exospheric density can be compared with the planetographic distributions of dust obtained by Mio/MDM and of the average particle precipitation obtained with the ion detectors of the two spacecraft.

# 5 Summary and Conclusions

With a full complement of in-situ and remote sensing instruments, BepiColombo can measure both the upstream conditions, magnetospheric and exospheric particles, and surface features during a solar event simultaneously. The availability of simultaneous two-point measurements will offer an unprecedented opportunity to investigate magnetospheric and exospheric dynamics at Mercury in response to Space Weather events during solar cycle 25, and to thereby investigate how these events provide and sustain different planetary responses. The possible contribution of other measurements of the Sun and solar wind conditions provided by other inner-heliosphere space missions will add a further contribution to the powerful science of BepiColombo.

The science of the environment is event-driven, so that, it is particularly important to use all the possible optimal configurations of the BepiColombo spacecraft, MPO and Mio, for being able to catch all specific events, such as ICME passage. The Hermean Environment Working Group of the BepiColombo mission has begun to identify the most important coordinated observations that will answer the crucial scientific questions to understand the functioning of Mercury's environment. Each single answer will constitute a step forward to answer the more general questions related to our Solar System formation and even more generally to the conditions of the exo-planetary systems. In fact, given the significant impact of stellar winds and radiation on exoplanets discovered close to their parent star, the investigation of the solar wind interaction with Mercury—the innermost planet in the Solar System—may have important immediate implications for studying exo-planetary conditions (e.g., Mura et al. 2011; Dong et al. 2017, 2018).

Assuming that we can depict the early-Sun conditions, the proposed coordinated observations, by clarifying the mechanisms responsible for the current net loss of planetary material, will contribute to estimate the past loss of Mercury. At the end, we could contribute answering to the long-standing question: Why is Mercury's density so high with respect to what is expected given its dimensions? Which are the historical contributions of the planetary mass loss caused by proximity to the Sun compared to that due to a possible giant impact?

Finally, the extreme and peculiar conditions of Mercury constitute a natural laboratory for investigating the kinetic regimes of a dynamic and small (compared to the ion gyroradius scales) magnetosphere, thus coordinated observations by BepiColombo will contribute to a better understanding of the mechanisms at the base of fundamental plasma physical processes.

# 6 Acronyms

| | |
|---|---|
| ASPERA | Analyzer of Space Plasma and EneRgetic Atoms |
| CIR | Corotating Interaction Regions |
| CME | Coronal Mass Ejection |
| ENA | Energetic Neutral Atom |
| ESD | Electron Stimulated Desorption |
| FIPS | Fast Imaging Plasma Spectrometer |
| FTE | Flux Transfer Events |
| GCR | Galactic Cosmic Ray |
| HSS | High Speed Streams |
| ICME | Interplanetary Coronal Mass Ejection |
| IMF | Interplanetary Magnetic Field |
| IPS | InterPlanetary Scintillation |
| KH | Kelvin-Helmholtz |
| LEP | Low-Energy Particles |
| LLBL | Low-Latitude Boundary Layer |
| $M_A$ | Alfvénic Mach Number |
| MESSENGER | Mercury Surface, Space Environment, Geochemistry and Ranging |
| MHD | MagnetoHydroDynamic |
| MASCS | Mercury Atmospheric And Surface Composition Spectrometer |
| MIV | Micrometeoroid Impact Vaporization |
| PIXE | Particle-Induced X-Ray Emission |
| PSD | Photo-Stimulated Desorption |
| SEP | Solar Energetic Particle |
| STEREO | Solar Terrestrial Relations Observatory |
| TAA | True Anomaly Angle |
| ULF | Ultra-Low Frequency |
| UV | UltraViolet |
| UVVS | UltraViolet and Visible Spectrometer |
| XRS | X-Ray Spectrometer |

**Acknowledgements** This work has been supported by the ASI-INAF agreement no. 2018-8-HH.O "Partecipazione scientifica alla missione BEPICOLOMBO SERENA Fase E1".